\documentclass{article}
\usepackage[a4paper,width=150mm,top=25mm,bottom=25mm,bindingoffset=6mm]{geometry}
\usepackage{xcolor}
\usepackage{graphicx}
\usepackage[utf8]{inputenc}
\usepackage[T1]{fontenc}
\usepackage[sorting=none]{biblatex}
\usepackage{mdframed}
\usepackage{framed}
\usepackage{amsfonts,amsmath,amsthm,amssymb,enumerate,bm}
\usepackage{tcolorbox}
\usepackage[pdfborder={0 0 0},pdfpagemode=UseOutlines,bookmarks=false,breaklinks=true,colorlinks=true,linkcolor=blue,citecolor=blue,urlcolor=blue]{hyperref}

\numberwithin{equation}{section}

\theoremstyle{plain} \newtheorem{theorem}{Theorem}[section]
\theoremstyle{plain} 
\theoremstyle{plain} 
\theoremstyle{plain} 
\theoremstyle{definition} 
\theoremstyle{definition} 
\theoremstyle{remark} 
\theoremstyle{remark} \newtheorem{algorithm}[theorem]{Algorithm}
\theoremstyle{remark} 

\def\R{\mathbb{R}}

\def\N{\mathbb{N}}
\def\Z{\mathbb{Z}}

\def\E{{\sf E}}
\def\P{{\sf P}}
\def\I{{\sf I}}

\def\e{{\rm e}}

\def\var{{\rm var}}

\begin{document}

\vspace*{10truemm}

\centerline{\sc \Large
A Unified Perspective on Exponential Tilt} 
\vspace{1mm}
\centerline{\sc \Large
and Bridge Algorithms
for Rare Trajectories}
\vspace{1mm}
\centerline{\sc \Large
of Discrete Markov Processes}

\vspace{5mm}

\centerline{{\small Javier Aguilar}$^{(1)}$ and {\small Riccardo Gatto}$^{(2)}$} 
\vspace{2mm}

\centerline{\small \it Submitted on July 24 2023, revised on September 11 2023 and on February 2 2024}
\vspace{4mm}

\centerline{\bf Abstract}
\noindent
This article analyzes and compares two general 
techniques of rare event simulation for generating paths of Markov 
processes over fixed time horizons: 
exponential tilting and stochastic bridge. 
These two methods allow to accurately
compute the probability that a Markov process ends within a rare region, 
which is unlikely to be attained. 
Exponential tilting is a general technique
for obtaining  an alternative or tilted sampling probability measure,
under which the Markov process becomes likely 
to hit the rare region at terminal time.
The stochastic bridge technique involves conditioning paths towards two endpoints: the terminal point and the initial one. The terminal point is generated from 
some appropriately chosen probability distribution that covers well 
the rare region. 
We show that both methods belong to the class of importance sampling 
procedures, by providing a common mathematical framework 
of these two conceptually different
methods of sampling rare trajectories.
\begin{color}{black}We also conduct a numerical comparison of these two methods, revealing distinct areas of application for each Monte Carlo method, where they exhibit superior efficiency.\end{color}
Detailed simulation algorithms are provided.
\vspace{2mm}

\centerline{\bf Key words and phrases}
\noindent
Backtracking --
Exponential tilting --
Importance sampling --
Likelihood ratio process -- 
Meta-stable state --
Monte Carlo simulation --
Radon-Nikodym derivative -- 
Rare event probability --
Relative error --
Stochastic bridge

\vspace{2mm}

{\footnotesize
\centerline{\bf Addresses} 
\noindent
1. {\it Corresponding author}. Investigador ForInDoc del Govern de les Illes Balears en el departamento de Electromagnetismo y Física de la Materia e Instituto Carlos I de Física Teórica y Computacional, Universidad de Granada, Granada E-18071, Spain -- 
{\tt javieraguilar@ifisc.uib-csic.es}
\\
Instituto de F\'isica Interdisciplinar y Sistemas Complejos IFISC (CSIC-UIB), Campus UIB, 07122 Palma de Mallorca, Spain.
\\
2. Institute of Mathematical Statistics and Actuarial Science, University of Bern,
Alpeneggstrasse 22, 3012 Bern, Switzerland --
{\tt riccardo.gatto@unibe.ch} --
 {\tt orcid.org/0000-0001-8374-6964}
\\\\
\centerline{\bf Acknowledgement}
The authors thank the anonymous Referees for constructive suggestions. They also thank
Raúl Toral, Pablo I. Hurtado, Carlos Pérez-Espigares, Rubén Hurtado,  Rapha\"el Chetrite and David Ginsbourger for useful discussions. \\
Research partially supported by the 
University of Bern,
UniBE ID Grant 2021, and the Conselleria d'Educaci\'o, Universitat i Recerca of the Balearic Islands (Grant FPI FPI\_006\_2020). Further financial support has been received from the Agencia Estatal de
Investigación (AEI, MCI, Spain) MCIN/AEI/10.13039/ 501100011033, Fondo
Europeo de Desarrollo Regional (FEDER, UE) under Project APASOS
(PID2021$\textbf{-}$ 122256NB$\textbf{-}$C21$/$C22), the María de Maeztu Program for units of
Excellence in $\text{R\&D}$, grant CEX2021$\text{-}$ 001164$\text{-}$M, and the contract ForInDoc (GOIB).}

\newpage

\tableofcontents
\vspace*{2truecm}


\newpage

\section{Introduction}

\noindent Rare trajectories account for realizations of stochastic processes that are extremely unlikely to happen, in the sense that conventional methods to sample stochastic trajectories, like the Euler-Maruyama methods in the context of stochastic differential equations~\cite{asmussen2007stochastic,toral2014stochastic} or the Gillespie algorithm for \begin{color}{black}pure\end{color} jumping processes~\cite{masuda2022gillespie,JAsampling_trajectories}, mostly fail to 
provide
satisfactory
statistical characterization within affordable times. 
Also, rare paths usually occur in particular shapes and unveil spatiotemporal patterns~\cite{hurtado2020building,ciccarese2022rare}.
Even if such paths are uncommon, they can have a decisive role in nature:
catastrophic events such as extinction of species~\cite{mobilia2010fixation,kessler2007extinction},
extreme rainfalls~\cite{frei2001detection} or earthquakes~\cite{gabrielov2000colliding} manifest themselves through rare fluctuations. 
\begin{color}{black}
The estimation of improbable events holds significance across various scientific domains. These include scenarios such as determining the probability for 
a medical therapy efficacy falling below a low threshold~\cite{ramoso_stochastic_2020},
the level of a dam exceeding a certain (very small or very large)
threshold~\cite{harrison1976stationary},
the capital of an insurance company falling below zero~\cite{gerber1979introduction} or a cosmic radiation corrupting memory cells in silicon microchips~\cite{1156060}.\end{color}
Therefore, the challenge of generating unlikely paths by using conventional methods, coupled with their significance in numerous phenomena,  motivates the need for sophisticated algorithms for sampling rare paths.

The literature contains numerous Monte Carlo algorithms that thwart the rarity of the event to simulate, thus making it possible to control the simulation error. 
The original reference in the statistical literature 
on importance sampling
by exponential tilting for first passage times of stochastic processes
is \cite{siegmund1976importance}; see also \cite{asmussen2007stochastic}, p. 164-166.
Some methods of rare event simulation rely on the generation of the bridge process, which conditions paths to the endpoints and associates rare trajectories to the occurrence or rare pairs of endpoints ~\cite{aguilar2022sampling,majumdar2015effective,wang2022brownian,wang2020exact,de2022optimal,orland2011generating,koehl2022sampling}. Other algorithms like cloning, also called splitting ~\cite{lecomte2007numerical,perez2019sampling,brewer2018efficient,grassberger2002go}, or like Metropolis schemes ~\cite{claussen2015convex,hartmann2020convex} link rare trajectories to the occurrence of rare sample averages of integrated quantities. 
A survey of simulation of rare events in queueing and reliability models can be found in \cite{10.1145/203091.203094}.
Following a theoretical approach, 
\cite{blanchet2012state} provides 
a review of central concepts of rare event simulation 
for light- and heavy-tailed systems. 
A recent presentation of importance sampling in the 
context of large deviations theory is given by
\cite{guyader2020efficient}.
Then \cite{e19110626} discusses variational formulations of the thermodynamic free energy within the framework of importance sampling and \cite{oakes2020deep}
 addresses the problem of automated search for optimal importance sampling
 schemes by using recent ideas from deep learning. 

Actually, the high variety of methods makes it difficult to assess which is the best strategy to tackle the generation of rare paths in specific problems. Moreover, it is not always simple to convey to what extent the
available Monte Carlo algorithms are fundamentally different or simply different expressions of the same mathematical framework.
The present article shows that two families of methods, those relying on exponential tilting and on bridge processes, can be understood within the common framework of importance sampling. This article focuses on their analytical and numerical comparison. These two methods are relevant as they encompass many strategies for sampling rare paths in non-equilibrium processes within finite time horizons. The present article does not tackle the theory of asymptotic optimality of estimation of rare event probabilities\footnote{The two usual criteria of asymptotic optimality of Monte Carlo estimators of rare event probabilities are
logarithmic efficiency and bounded relative error, for which
we refer to p. 158-160 of \cite{asmussen2007stochastic} or to \cite{doi:https://doi.org/10.1002/9781118445112.stat07823}.\label{fo2}},
which relies on the theory of large deviations; see e.g. \cite{dembo1998large} for a general reference. 
Neither it covers
methods that do not use changes-of-measure, 
such as the forward flux sampling~\cite{allen2009forward,hall2022practical}, the transition path sampling~\cite{van2012dynamical,bolhuis2010trajectory,dellago2009transition} or the saddlepoint technique, which provides the deterministic most likely path in the limit of small noise~\cite{heymann2008pathways,grafke2019numerical}.

More precisely, the scope of this article is the following.
The first and main objective is to show that these
two conceptually different techniques
can be re-expressed in terms of the same change-of-measure formula
for the expected value. Thus,
the two algorithms differ only in their likelihood ratio
viz. Radon-Nikodym derivative, which is the factor
that accounts for the replacement of the original
sampling probability measure by a new one.
The second objective
of this work is to compare the numerical performance of these two techniques in various settings. We provide numerical evidence that 
in real situations with finite time horizons, 
the relative errors of backtracking and exponential tilting 
appear bounded. This provides a surrogate for theoretical results of asymptotic optimality. For both methods,
we numerically show that there is an optimal parameter minimizing 
the relative error. Finally, we find that stochastic bridges provide superior suitability for tackling challenges associated with rare transitions between meta-stable states.

As a byproduct of demonstrating the precise emergence of these two methods from the technique of change-of-measure, we establish that stochastic bridge techniques may introduce systematic errors in the estimation of averages, which is a novel theoretical result within the field. Furthermore, we obtain a fairly complete tutorial on these two techniques of rare event simulation, since we offer readily applicable algorithms that streamline the implementation of these concepts into computer programs.
For the sake of simplicity, many measure-theoretic details are given separately in footnotes.

The rest of the article has the following structure.
Section~\ref{s2}
reviews the general theory of change-of-measure
for stochastic processes and in particular for
Markov processes with discrete time and state spaces.
In Sections~\ref{s3} and~\ref{sec:backtracking} we recast
exponential tilting and bridge change-of-measure. 
Numerical applications
of these two Monte Carlo techniques to
a simple binomial process and to a process with meta-stable states
are provided in Section~\ref{s4}.
Concluding remarks are presented in Section \ref{s5}. 
Supplemental material (SM)~\cite{SM}, Section \ref{algo},
shows explicitly the Monte Carlo 
algorithms introduced in this article and
SM~\cite{SM}.

\section{Change-of-measure and likelihood ratio process}    \label{s2}
This section provides a succinct introduction to the theory of 
change-of-measure. We refer the interested reader to \cite{asmussen2007stochastic,doi:https://doi.org/10.1002/9781118445112.stat07823,bucklew2004introduction,asmussen2003applied} for a more in-depth introduction to the topic.
We start with the case of the single random variable
and we then generalize this to the stochastic process.

\subsection{Random variable}\label{sec:RV}

One of the central ideas in rare event simulation is the change of the sampling measure, which allows transforming the problem of estimating averages over some probability measure $\P$ to another average estimation over a different measure $\tilde{\P}$. The idea is that suitable transformations of this kind can ease the estimation of averages. Let us present the basic result of the theory. Let $(\Omega, {\cal F},\tilde{\P})$ 
be a probability space~\cite{paul1999stochastic}.
Assume that the random variable $L$ over this space is
nonnegative with $\tilde{\P}$ probability
1, i.e. $\tilde{\P}$-almost surely ($\tilde{\P}$-a.s.),
and satisfies 
$\E_{\tilde{\P}}[L] = 1$. Then, one shows that
\begin{equation}                               \label{e391}
{\P}[A] = \E_{\P} [ I_A ]= \E_{\tilde{\P}} [ I_A L ] = \int_A L d \tilde{\P}, \; \forall A \in {\cal F},
\end{equation}
defines an unique probability measure ${\P}$ on $(\Omega, {\cal F})$. 
In Eq.~\eqref{e391} we denote the indicator as $$I_A(\omega)=I\{\omega\in A\}=\begin{cases}
    1, \quad \text{if  }\omega\in A, \\
    0, \quad \text{otherwise.}
\end{cases}$$
Therefore, computing the probability $\P[A]$, which is the expectation of $I_A$ under the measure $\P$, is equivalent to computing the expectation, this time under $\tilde{\P}$, of 
$I_A L$. The above result is a fairly general one, since the only restriction on the new measure $\Tilde{\P}$ is absolute continuity with respect to
(w.r.t.) ${\P}$, denoted $\P \ll \tilde{\P}$ on ${\cal F}$.
This means $\tilde{\P}[A] = 0 \Rightarrow 
{\P}[A]=0$, $\forall A \in \cal{F}$. In other words, any set $A$ allowed by ${\P}$ must be allowed
by $\tilde{\P}$ as well.
The random variable $L$, often called likelihood ratio, is the 
Radon-Nikodym derivative of ${\P}$ w.r.t. $\tilde{\P}$, denoted
$d {\P} / d \tilde{\P}$. 

Eq. (\ref{e391}) gives the following 
change-of-measure result.
Let $X$ be a random variable on $(\Omega, {\cal F})$, then
\begin{align}                         \label{e392}
z=\E_{{\P}}[X] = \E_{\tilde{\P}}[ X L],
\end{align}
when ${\P} \ll \tilde{\P}$.\footnote{Rigorously,
Eq. (\ref{e392}) requires ${\P} \ll \tilde{\P}$ only on the restriction of ${\cal F}$ to 
$\sigma(X) \cap \{X \neq 0\}$, where
$\sigma(X)=\{X^{-1}(B)| B \in {\cal B}(\R)\}$
is the $\sigma$-algebra generated by $X$.}

A simple illustration is the following. Let $X$ be 
Gaussian with mean 0 and variance equal to 1, under $\P$, let 
$\mu \in \R$ and let
\begin{align}                           \label{e2000}
L=\exp\left\{-\mu (X-\mu) - \frac{\mu^2}{2} \right\}. 
\end{align}
We have from Eq. (\ref{e392}) that
$$\E_{\P}\left[\e^{v X}\right] = 
\E_{\tilde{\P}}\left[\e^{v X}L\right]=
\e^{\frac{1}{2}v^2}, \; \forall v \in \R,$$
iff $X$ is Gaussian with mean $\mu$ and variance equal to one under
$\tilde{\P}$. So this change-of-measure allows for arbitrary 
recentering of $X$, yet without changing the variance of $X$. 

Let $f$ and $\tilde{f}$ be the densities of the random variable $X$ 
under $\P$ and $\tilde{\P}$, respectively.
We then have that 
\begin{equation*}                                               
L = \frac{f(X)}{\tilde{f}(X)}
\end{equation*}
is a valid likelihood ratio for the change-of-measure
$$
z = \E_\P[g(X)]=
\E_{\tilde{\P}} [g(X) L] = \int_\R g(x) \frac{f(x)}{\tilde{f}(x)} \tilde{\P} 
[ X \in ( x , x + d x) ],\footnote{Precisely, the validity of 
$L=f(X)/\tilde{f}(X)$ 
is limited to
the restriction of ${\cal F}$ to $\sigma(X)$, meaning that
$\E_\P[Z]= \E_{\tilde{\P}} [Z L]$ would be untrue 
with $Z$ not $\sigma(X)$-measurable, viz. with any $Z$ that 
could not take the 
form $Z=g(X)$, for some Borel function $g$.}$$
for any Borel $g\!:\R \to \R$.
In this situation,
$ \P \ll \tilde{\P}$ can be re-expressed as the support of 
the density $f$ 
being included into the support of the density $\tilde{f}$.
We see directly that the likelihood ratio in Eq. (\ref{e2000}) 
is indeed the ratio of the two given Gaussian densities, evaluated
at $X$.

The importance sampling algorithm amounts to select a large number of replication $m$, to
generate $X_1 , \ldots, X_m $ independently from $\tilde{f}$ and then to
estimate $z$ by
$$\hat{z}_m = \frac{1}{m} \sum_{j=1}^m g(X_j) 
\frac{f(X_j)}{\tilde{f}(X_j)}. $$

\subsection{Stochastic process}

In this section, we show the extension of Section~\ref{sec:RV} to the case of stochastic processes. Let $(\Omega, {\cal F},  \{ {\cal F}_t\}_{t \ge 0},\tilde{\P})$ be a
filtered probability 
space,\footnote{The sequence of $\sigma$-algebras
 $\{ {\cal F}_t\}_{t \ge 0}$ is a filtration in the
 sense that ${\cal F}_s \subset {\cal F}_t \subset {\cal F}$,
 $\forall \ 0 \le s \le t < \infty$ (with inclusion weakly meant).} where  time is either discrete, $t \in \N$, $\N=
 \{0,1,\ldots\}$, or continuous, $t \in [0,\infty)$.
Assume that the stochastic process $\{ L_t \}_{t \ge 0}$ 
over this space
is a
$\tilde{\P}$-a.s. nonnegative martingale
w.r.t. the filtration $\{{\cal F}_t \}_{t \ge 0}$\footnote{This means
that $\E_{\tilde{\P}}[ L_t | {\cal F}_s] = L_s$, $\forall \ 0 \le s \le t < \infty$.} 
such that $\E_{\tilde{\P}}[L_t] = 1$, $\forall t \ge 0$.
Then there exists a unique probability measure $\P$ on  $(\Omega, {\cal F})$ such that,
\begin{equation}                                                                        \label{e101}
\forall t \ge 0, \; {\P}[A_t] = 
\E_{\tilde{\P}}[ I_{A_t} L_t] = 
\int_{A_t} L_t d \tilde{\P}, \; \forall A_t \in {\cal F}_t .
\end{equation}
Thus $\P \ll \tilde{\P}$.\footnote{Precisely,
$\P \ll \tilde{\P}$ holds on the restriction 
of ${\cal F}$ to 
${\cal F}_t$, $\forall t \ge 0$.}
The martingale $\{L_t\}_{t \ge 0}$ is called Radon-Nikodym or likelihood ratio process. At any $t\ge0$,
$L_t$ is the density or Radon-Nikodym derivative
of ${\P}$ w.r.t. $\tilde{\P}$ on ${\cal F}_t$.\footnote{An alternative commonly used notation
is $L_t = d {\P} / d \tilde{\P} \! \mid_t$.} 
Proof of Eq. (\ref{e101}) can be found e.g. in \cite{asmussen2003applied}. Thus
Eq. (\ref{e101}) generalizes Eq. (\ref{e391}). We 
have the following change-of-measure result for stochastic 
processes:
for any integrable process
$\{ X_t \}_{t \ge 0}$,\footnote{The process $\{ X_t \}_{t \ge 0}$ must also be $\{ {\cal F}_t\}_{t \ge 0}$-adapted, in the sense that $X_t$ is ${\cal F}_t$-measurable, $\forall t \ge 0$.}
it holds that
\begin{align}                               \label{e394}
\E_{\P}[X_s] =
\E_{\tilde{\P}}[X_s L_s] =
\E_{\tilde{\P}}[X_s L_t],
\end{align}
provided 
$\P \ll \tilde{\P}$.\footnote{Precisely, $\P \ll \tilde{\P}$ is required on the restriction of
${\cal F}$ to
${\cal F}_s \cap \{ X_s \neq 0 \}$, $\forall 0 \le s \le t$.}

\subsection{Discrete Markov process}\label{sec:Backtracking_dt_ds}

Through the rest of the text, we will consider the Markov process with discrete time domain 
$\N$ and discrete state space $\Z=\{\ldots,-1,0,1,\ldots\}$. We note that most applied problems can be indeed formulated in this setting through time and space discretization, so this choice should not entail practical restrictions.
We obtain the likelihood ratio process of change-of-measure
from the induced probability of the Markov process. 
The likelihood ratio takes a simple form, depending only on the 
transition kernels of the Markov process. In this section
we consider the time $t \ge 1$ and the states
$n,n' \in \Z$.

Our Markov process $\{ X_t \}_{t \in \N}$ is defined on
the filtered probability space
 $(\Omega,{\cal F}, \{{\cal F}_t \}_{t \ge 0} ,\P)$.
Let us define the transition probabilities
\begin{align*}
p_{t,n}(j) & = \P [X_t = n + j \mid X_{t-1} = n], \;\;
\forall j\in \Z,
\end{align*}
together with the probabilities of the initial state
\begin{align*}
p_0(n) & =
\P [X_0 = n ].
\end{align*}

Let $\tilde{\P}$ denote a second probability measure on
$(\Omega,{\cal F}, \{{\cal F}_t \}_{t \ge 0})$, which is
unambiguously determined 
through the change-of-measure kernels 
$q_0\!: \Z \to [0,\infty)$ and
$q_t\!: \Z\times\Z \to [0,\infty)$ as follows:
\begin{align}                                   \label{e130} 
p_0(n) & = 
q_0(n) \tilde{\P} [X_0 = n]
\end{align}
and
\begin{align}                                   \label{e31}
{\P} [X_t = n' & \mid X_{t-1} = n, X_{t-2} = n_{t-2},\ldots, X_1 = n_1] = 
\nonumber \\
&  q_t(n,n') \tilde{\P} [X_t = n' \mid X_{t-1} = n, X_{t-2} = n_{t-2},
\ldots, X_1 = n_1],
\end{align}
for $n_{t-2}, \ldots,n_1 \in \Z$.
Because $\{ X_t \}_{t \in \N}$ is a Markov process under $\P$,
the values of $n_{t-2}, \ldots,n_1$ on the
left side of Eq. (\ref{e31}) are irrelevant. They remain
irrelevant on the right side and thus
$\{ X_t \}_{t \in \N}$
remains a Markov process
under the new probability measure $\tilde{\P}$. So 
we can simplify Eq. (\ref{e31}) to
\begin{align}                               \label{e110}
p_{t,n}(n'-n) & = 
q_t(n,n') \tilde{\P} [X_t = n' \mid X_{t-1} = n].
\end{align}
We can define the transition probabilities under $\tilde{\P}$ by
\begin{align}\label{eq:new_transition_prob}
\tilde{p}_{t,n}(j) & =
\tilde{\P} [X_t = n + j \mid X_{t-1} = n]=\frac{p_{t,n}(j)}{q_t(n,n+j)}, \; 
\forall j\in \Z,
\end{align}
and the initial probability under $\tilde{\P}$ by
\begin{align*}
\tilde{p}_0(n) & =
\tilde{\P} [X_0 = n ].
\end{align*}
We then have,
for $n_{0}, \ldots,n_m \in \Z$,
\begin{align*}
\frac{{p}_0 (n_0)}
{\tilde{p}_0(n_0)}
\prod_{t=1}^{m}
\frac{{p}_{t,n_{t-1}}(n_t-n_{t-1})}
{\tilde{p}_{t,n_{t-1}}(n_t-n_{t-1})} =
q_0(n_0) \prod_{t=1}^{m} q_t (n_{t-1},n_t), \; \text{ for }
m=1,2,\ldots.
\end{align*}
This last expression gives us the following general form of the likelihood ratio process,
\begin{align*}                                       
L_0 & = q_0 (X_0) \; \text{ and } \;
L_{m}  =
q_0(X_0) \prod_{t=1}^{m} q_t (X_{t-1},X_t), \; 
\text{ for }
m=1,2,\ldots.
\end{align*}
Thus $L_m$ is a function of $X_0,\ldots,X_m$, 
for $m=0,1,\ldots$.\footnote{In other terms, the 
likelihood ratio process is adapted to the 
filtration generated by the Markov process.}

When the Markov process is homogeneous under $\P$ and
the change-of-measure kernel $q_t$ does not depend on $t \ge 1$, then
the Markov process $\{ X_t \}_{t \in \N}$
remains homogeneous under
$\tilde{\P}$. In this case, 
by redenoting the change-of-measure kernels at times $t\neq 0$
in Eq. (\ref{e31})
simply by $q_\bullet$, we
obtain the likelihood ratio process 
\begin{align}                                   \label{e27}
L_0 & = q_0 (X_0) \; \text{ and } 
L_{m}   =
q_0(X_0) \prod_{t=1}^{m} q_\bullet (X_{t-1},X_t), \text{ for }
m=1,2,\ldots.
\end{align}

Thus the following change-of-measure formula holds for all 
events depending on the Markov process over the 
finite time horizon $[0,\ldots,t^{\dag}]$, for some $t^{\dag} \ge 1$. 
For some given function
$g\! :\Z^{t^{\dag}+1} \to \R$,
define the importance sampling estimator
$Z_{t^\dag}=g(X_0,\ldots,X_{t^{\dag}}) 
L_{t^{\dag}}$. We then have
\begin{align}                                   \label{e49}
z_{t^\dag}=&\E_{{\P}} [ g(X_0,\ldots,X_{t^{\dag}}) ]  =
\E_{\tilde{\P}} \left[ g(X_0,\ldots,X_{t^{\dag}}) 
L_{t^{\dag}}  \right] 
= \E_{\tilde{\P}} \left[ Z_{t^\dag} \right],
\end{align}
whenever $\P \ll \tilde{\P}$.\footnote{Precisely,
$\P \ll \tilde{\P}$ is required
on ${\cal F}_{t^\dag}$.}

Until now, nothing has been said about the form of the change-of-measure kernels ($q$). We will focus on two particular choices for these kernels, namely, the exponential tilt and bridge changes-of-measure.

\subsection{Absolute continuity and simulation}  \label{sec:absolute_continuity}


In the context of simulation, ${\P}$ represents the original measure 
and replications of $XL$, see Eq. (\ref{e392}), or of $X L_s$,
see Eq. (\ref{e394}),
are drawn under the importance sampling measure $\tilde{\P}$.
It may appear weird to state the existence of the original measure
$\P$ (which we already have) through 
Eq. (\ref{e391}) and Eq. (\ref{e101}), but the important
aspect here is the unambiguous
relationship between $\P$ and $\tilde{\P}$:
if either
Eq. (\ref{e391}) or Eq. (\ref{e101}) can be established, then the
importance sampling algorithm with $\tilde{\P}$ is valid.

From the theoretical perspective,
the only restriction for choosing the importance sampling measure 
($\tilde{\P}$)
is absolute continuity ($\P\ll\tilde{\P}$), which $\tilde{\P}$-a.s. 
guarantees the existence of the likelihood process ($\{L_t\}_{t \ge 0}$). 
A sample path with probability zero under the original measure 
($\P$) may thus receive positive probability under the importance
sampling measure ($\tilde{\P}$). However, since paths can be important observables themselves, one can be interested in a new measure 
($\tilde{\P}$)
that sample only the paths that have positive
probability under the original probability ($\P$).  
For example, in the context of stochastic thermodynamics, random paths have a prominent role in the characterization of entropy production~\cite{van2013stochastic,seifert2012stochastic}. As another example, rare paths can be measurable objects with important biological implications~\cite{ciccarese2022rare,wang2011quantifying}. Thus, it can also be useful to have the stronger constraint of equivalence of measures
($\P\ll\tilde{\P}$ and $\tilde{\P}\ll\P$).\footnote{Precisely, one assumes
$\P\ll\tilde{\P}$ and $\tilde{\P}\ll\P$ on
${\cal F}_t$, $\forall t \ge 0$.}
In Sections \ref{s3} we will see that the exponential tilting always satisfies this condition. Whereas in the \begin{color}{black} bridge \end{color} process, the equivalence of measures depends on the choice of the terminal distribution (see Section \ref{sec:backtracking}).

Lastly, we note that when only absolute continuity of the new measure w.r.t. the original one ($\tilde{\P}\ll\P$) holds, 
an importance sampling algorithm may still be used but
introduce systematic errors (also called bias errors) in the estimation of the quantity of interest $z$, given in Eq.~\eqref{e49}. 
Such systematic errors can be small, relative to the Monte
Carlo variability, if the forbidden region by $\tilde{\P}$ is irrelevant for the estimation of $Z$.
\begin{color}{black} Nevertheless, these systematic errors can also hold significance as they have the potential to surpass Monte Carlo (statistical) errors, a point we will illustrate through a numerical example in Section~\ref{sec:homogeneous_binomial_process}.
\end{color}

\section{Exponential tilt
for Markov processes}   \label{s3}

This section provides the analytical
formulation of importance sampling by exponential tilting.
The technique is first introduced for a single random variable,
then for the simple process of partial sums of i.i.d. random variables
and finally for discrete Markov processes. 

Exponential tilting is a fairly general change-of-measure
procedure that can be applied whenever the underlying distribution decays sufficiently fast at its extremities;
namely when the distribution is ``light-tailed''.
This procedure embeds the original probability measure into a 
new one, which renders likely specific
trajectories that would have been otherwise 
rare, under the original probability. This technique is sometimes called Esscher transformation.
It was indeed suggested by
\cite{escher1932probability,esscher1963approximate}
for local applications of the central limit theorem,
in order to obtain a very accurate analytical approximation
to the distribution of the sum. 
It was then shown by \cite{daniels1954saddlepoint} that
these Esscher's approximation can be reformulated in terms the saddlepoint
approximation of asymptotic analysis~\cite{copson2004asymptotic}.
Theoretically, both saddlepoint approximation and optimal
exponential tilting belong to the class of
large deviations
approximations~\cite{gatto2014saddlepoint}.
These approximations are adequate
for obtaining the very small probabilities of rare events;
see e.g. Chapter 3 of~\cite{bucklew2004introduction}.

Exponential tilting is introduced in
Section \ref{s32} for a single random variable. It is
then given for the sum of i.i.d. random variables,
in Section \ref{s33}.
The likelihood ratio process of
exponential tilt for Markov processes is provided in 
Section \ref{s34}. 
We conclude with two remarks in Section \ref{s30}:
Section \ref{s35} concerns the choice of the tilting
parameter and Section \ref{sec:s-ensemble}
presents a closely related method, called
$s$-ensemble.

\subsection{Exponential tilt for random variable}           \label{s32}

Consider now the random variable $X$ with 
cumulant generating function (c.g.f.)
$$K(\theta) = \log \E_\P \left[ \e^{\theta X} \right], \quad \text{for } \theta \in \R,$$
\begin{color}{black} where $\P$ represents the present probability measure, and we consider values of $\theta$ such that $K(\theta)$ is finite.
\end{color}
The exponentially tilted measure $\P_{\theta}$ 
is the measure $\tilde{\P}$ of Eq. (\ref{e101}) obtained by
Radon-Nikodym derivative or
likelihood ratio 
\begin{equation}                      \label{e109}                   
L_\theta = 
\frac{d \P}{d \P_{\theta}} = \exp \{ - \theta X + K(\theta)\},
\end{equation} 
over
$\sigma(X)$. 
Thus $\P_\theta$
is equivalent to $\P$, 
in the sense that $\P_\theta \ll \P$ and 
$\P \ll \P_\theta$.\footnote{Precisely, 
$\P_\theta \ll \P$ and 
$\P \ll \P_\theta$ hold 
on on the restriction of ${\cal F}$ to $\sigma(X)$.} The new measure
$\P_\theta$, called exponential tilt of $\P$, is a
practical importance sampling measure. We note that
if $F$ is the distribution function (d.f.) of $X$
under $\P$, then
\begin{align}                                   \label{e88}
 d F_{\theta} (y) = \exp\{ \theta y - K(\theta) \}
d F(y)
\end{align}
provides the d.f. under $\P_\theta$.

Although our focus lies on univariate processes, we can briefly
mention that exponential tilting generalizes directly to 
the multivariate setting. 
When $\bm{X}$ is a random vector in $\R^d$, for some $d \ge 2$, 
with c.g.f. 
$K(\bm{v}) = \log \E_\P \left[ \e^{\langle \bm{v},\bm{X} \rangle} \right]$, 
for $\bm{v} \in \R^d$, then the likelihood ratio of Eq. (\ref{e109}) becomes
\begin{equation}                                    \label{e30}                                    
L_{\bm{\theta}} = 
\frac{d \P}{d \P_{\bm{\theta}}} = \exp \{ - \langle \bm{\theta} , \bm{X} \rangle + K(\bm{\theta})\},
\end{equation} 
for $\bm{\theta} \in \R^d$.

\subsection{Exponential tilt for random walk}                                        \label{s33}

Let us introduce the exponential tilt likelihood ratio
process for the simple random walk, which is the process of partial sums of independent random
variables $Y_1, Y_2, \ldots$ with common d.f. $F$ and c.g.f. $K$, 
under some probability measure $\P$. Consider thus
$$X_t = \sum_{j=1}^t Y_j, \text{ for } t=1,2,\ldots.$$ 
We can assume the fixed initial 
state $0$, viz. define $X_0=0$.
The exponentially tilted measure $\P_{\theta}$ over
$\sigma(X_1,\ldots,X_t)$ is
obtained from the likelihood ratio process
\begin{align*}
L_{t}(\theta) = \exp \{ - \theta X_t  + t K(\theta) \},
\;\; \mbox{for} \; t=1,2\ldots,
\end{align*}
with $\theta$ such that $K(\theta)$ is finite.
We have $\P \ll \P_\theta$ and $\P_\theta \ll \P$.\footnote{Precisely,
$\P_\theta \ll \P$ and $\P \ll \P_\theta$ hold
on $\sigma(X_1,\ldots,X_t)$, 
which is the $\sigma$-algebra generated by 
$X_1,\ldots,X_t$,
for $t=1,2,\ldots$.}

For a given time horizon $t^\dag \ge 1$ and for a given 
function
$g\! :\Z^{t^{\dag}} \to \R$, we are generally interested
in computing
$z_{t^\dag}=\E_{\P} [ g(X_1,\ldots,X_{t^{\dag}}) ]$.
The importance sampling estimator
of exponential tilting is given by
\begin{align}                                   \label{e500}     
Z_{t^\dag}(\theta)=g(X_1,\ldots,X_{t^{\dag}}) 
L_{t^{\dag}}(\theta)
\end{align}
and we have
\begin{align}                \label{e79}                           
z_{t^\dag}=&\E_{{\P}} [ g(X_1,\ldots,X_{t^{\dag}}) ]  =
\E_{\P_{\theta}} \left[ g(X_1,\ldots,X_{t^{\dag}}) 
L_{t^{\dag}}(\theta) \right] 
= \E_{\P_{\theta}} \left[ Z_{t^\dag}(\theta) \right].
\end{align}

Given the multidimensional likelihood ratio formula of Eq. (\ref{e30}), the generalization of 
the above one-dimensional exponential tilting to the
random walk $\{\bm{X}_t\}_{t \ge 0}$
with individual values 
in $\R^d$ with $d \ge 2$ is straightforward.

Note that sampling under $\P_{\theta}$ amounts to
generate i.i.d. summands 
from the exponentially tilted d.f. of Eq. (\ref{e88}).
We thus obtain Algorithm \ref{al020} of SM~\cite{SM}, Section \ref{algo}, for importance sampling
by exponential tilt for random walks.

For some large 
$x > \E_\P[ Y_1 ]$, let
$\I_x= (t^\dag x , \infty)$, for some time horizon $t^\dag \ge 1$.
We are now interested in the rare event probability 
$z_{t^\dag}(\I_x) = \P [ X_{t^\dag}\in \I_x ]$, which a small upper
tail probability of the sample mean.
The importance sampling estimator is thus given by
\begin{align}                               \label{e19}
Z_{t^\dag}(\theta,\I_x) = I\{ X_{t^\dag} \in \I_x \} L_{t^\dag}(\theta)= I\{ X_{t^\dag} > t^\dag x\} L_{t^\dag}(\theta)
\end{align}
and we have
$z_{t^\dag}(\I_x) = \E_{\P_\theta}[Z_{t^\dag}(\theta,\I_x)]$.

But not every choice of tilting parameter $\theta$ reduces the
variability and inadequate choices may also increase it, substantially.
Let $\theta(x)$ the solution w.r.t. $v$ 
of 
\begin{align}                                       \label{e47}
K'(v) = \frac{d}{d v} K(v) = x
\; \text{ i.e. } \;
\E_{\P_{\theta(x)}}[Y_1]=x
\; \text{ i.e. } \;
\E_{\P_{\theta(x)}}[X_{t^\dag}]=t^\dag x.
\end{align}
It is shown that $\theta(x)$ exists and it is unique, for any
$x$ within the interior of the range of $K'$; see e.g.
\cite{daniels1954saddlepoint}.
Moreover, \cite{daniels1954saddlepoint} shows also
that $\theta(x)$ appears as a saddlepoint
on the surface of the real part of the complex exponent
of the Fourier transform of the density. 
It is shown 
at p. 168-169 of \cite{asmussen2007stochastic} that
the importance sampling estimator given in Eq. (\ref{e19}) with
$\theta = \theta(x)$
is optimal, in the sense of logarithmic efficiency,
under $\P_{\theta(x)}$.
A less rigourous but simple justification
of the optimality of this choice of titling parameter
is given at Section \ref{s35}.

Logarithmic efficiency is a slightly weaker criterion 
than bounded relative error, see Footnote \ref{fo2}.
These two usual optimality criteria of rare event simulation
are asymptotic for vanishing probabilities like $z_{t^\dag}(\I_x)$,
as $x\to \infty$.
For many important accurate estimators of small probabilities, 
only logarithmic efficiency can be established~\cite{asmussen2007stochastic}. 
However, these two criteria can hardly be
distinguished in most practical situations.

A detailed presentation of importance sampling in the 
context of large deviations theory is given by
\cite{guyader2020efficient}. In particular,
an analysis of the joint large deviations behaviour of
the random process of interest (called ``observable'')
and of the (logarithmically rescaled) likelihood
ratio process is presented.
The article 
provides necessary and sufficient 
conditions, for
any general change-of-measure procedure (not necessarily exponential tilt), in order to have logarithmic efficiency.
Interestingly, this result motivates further research concerning
the existence and the characterization of logarithmically efficient change-of-measure methods other than exponential tilt.

Note finally the following property: the exponentially tilted 
distribution $\P_{\theta(x)}$ is the closest one to
the original distribution $\P$, under all
distributions that are centered according to Eq.~(\ref{e47}), where
closeness is
in terms of Kullback-Leibler divergence; cf. e.g. \cite{gatto2014saddlepoint}.

\subsection{Exponential tilt for homogeneous Markov processes}           \label{s34}

The discrete time Markov process, its change-of-measure
kernels and and its likelihood ratio process are all introduced 
in Section \ref{sec:Backtracking_dt_ds}. 
We showed that the change-of-measure kernels allow us to
obtain an alternative probability measure $\tilde{\P}$.
The objective is to choose $\tilde{\P}$ so to
reorient sample paths towards a specific region of interest, which 
is rarely reached under
the original measure $\P$. We show here how to $\tilde{\P}$
is obtained through exponential tilt, for 
the homogeneous Markov process. 
We only need to obtain the change-of-measure kernels
of Eq. (\ref{e130}) and Eq. (\ref{e31}) of exponential tilting.
In this section
we consider the time $t \ge 1$ and the states
$n,n' \in \Z$.

\noindent Let 
\begin{align}                               \label{e84}
z(n,\theta) & = 
\exp \{ \theta n - K_{0}(\theta)\}
\end{align}
and
\begin{align}                               \label{e94}
z(n,n',\theta) & = 
\exp \{ \theta (n'-n) - K_{\bullet n}(\theta)\}, 
\end{align}
where 
\begin{align}                                   \label{e158}
K_{0}(\theta) = \log \sum_{j \in \Z} \exp\{\theta j\} 
p_{0}(j) 
\end{align}
and
\begin{align}                                   \label{e159}
K_{\bullet n}(\theta) = \log \sum_{j \in \Z} \exp\{\theta j\} 
p_{\bullet n}(j),
\end{align}
at any $\theta \in \R$ where the two sums above converge,
are the c.g.f.  
of the probability of the initial state, denoted $p_0$, and
the c.g.f.
of the homogeneous
transition probabilities, denoted $p_{\bullet n} = p_{t, n}$ and
independent of the time index $t\ge 1$.
The exponentially tilted probability measure,
$\tilde{\P} = \P_\theta$, is characterized by
\begin{align}                                           \label{e38}
 p_{0}(n,\theta) & =  \P_\theta [X_0 = n ] 
  = \exp \{ \theta n - K_{0}(\theta)\} 
     \P [X_0 = n] 
= z(n,\theta) p_{0}(n)
\end{align}
and
\begin{align}                                           \label{e40}
 p_{\bullet n}(j,\theta) & =  \P_\theta [X_t = n + j \mid X_{t-1} = n]  = \exp \{ \theta j - K_{\bullet n}(\theta)\} 
     \P [X_t = n + j \mid X_{t-1} = n] \nonumber \\
& = z(n,n+j,\theta) p_{\bullet n}(j), \; 
\forall j\in \Z.
\end{align}

Thus exponential tilt corresponds to
the particular choice of the general change-of-measure kernels
Eq. (\ref{e130}) and Eq. (\ref{e31}), respectively given by
$$
q_0(n) = \frac{1}{z(n,\theta)}
\; \text{ and } \;
q_\bullet(n,n') = \frac{1}{z(n,n',\theta)}.$$ 

Thus, the likelihood ratio process of Eq. (\ref{e27})
for the case of exponential tilt
becomes
\begin{align}                                   \label{e29}
L_0(\theta) & = q_0 (X_0) =[z(X_0,\theta)]^{-1}\; \text{ and } \nonumber \\
L_{t}(\theta) & = 
\left[ z(X_0,\theta) \prod_{k=1}^{t} z(X_{k-1},X_k,
\theta) \right]^{-1} 
\nonumber \\
& = \left[ \exp \left\{ \theta X_0 - K_0(\theta)  +
\sum_{k=1}^{t} \theta (X_k - X_{k-1}) - K_{\bullet X_{k-1}} 
(\theta)\right\} \right]^{-1} \nonumber \\
& = \exp \left\{ - \theta X_t + \left[ K_0(\theta)  +
\sum_{k=1}^{t} K_{\bullet X_{k-1}} 
(\theta)\right] \right\} \nonumber \\
& = \e^{-\theta X_t} M_0(\theta)
\prod_{k=1}^{t} M_{\bullet X_{k-1}} (\theta), 
\end{align}
where the argument $\theta$ has been added to the likelihood ratio 
for convenience and where $M_{0} = \e^{K_{0}}$ and
$M_{\bullet n} = \e^{K_{\bullet n}}$ are
the moment generating functions of 
$p_0$ and $p_{\bullet n}$, 
respectively.

Let us now give a couple of remarks.
The required absolute continuity is clearly satisfied,
because of the positivity of the change-of-measure kernels of exponential tilt.
In fact, both ${\P} \ll {\P}_\theta$ and
${\P}_\theta \ll {\P}$ hold, viz. 
${\P}$ and ${\P}_\theta$ are equivalent.
In contrast with the likelihood ratio for the bridge process, presented in Section~\ref{sec:backtracking}, the likelihood ratio of Eq. (\ref{e29}) 
is not restricted to problems with finite time horizons.

Consider the time horizon $t^\dag \ge 1$ and the  
function
$g\!: \Z^{t^{\dag}+1} \to \R$. We want to evaluate
$               
z_{t^\dag} = \E_{{\P}} [ g(X_0,\ldots,X_{t^{\dag}}) ]$.
The estimator of exponential tilting is given by
\begin{align}                                   \label{e344}
Z_{t^\dag}(\theta)=g(X_0,\ldots,X_{t^{\dag}}) 
L_{t^{\dag}}(\theta),
\end{align}
for $L_{t^{\dag}}(\theta)$ given in Eq. (\ref{e29}),
and we have
\begin{align}                                     \label{e129}   
z_{t^\dag} = \E_{{\P}} [ g(X_0,\ldots,X_{t^{\dag}}) ]  =
\E_{\P_{\theta}} \left[ g(X_0,\ldots,X_{t^{\dag}}) 
L_{t^{\dag}}(\theta)  \right] 
= \E_{\P_{\theta}} \left[ Z_{t^\dag} (\theta)\right].
\end{align}

Exponential tilting can be also obtained for the multidimensional
homogeneous Markov process
$\{\bm{X}_t\}_{t \ge 0}$
taking individual 
values in $\Z^d$, at any $d \ge 2$, essentially by replacing Eq. (\ref{e84}) and Eq. (\ref{e94})
by
\begin{align*}
z(\bm{n},\bm{\theta}) & = 
\exp \{ \langle \bm{\theta}, \bm{n} \rangle - K_{0}(\bm{\theta})\}
\; \text{ and } \;                       
z(\bm{n},\bm{n}',\bm{\theta}) = 
\exp \{ \langle \bm{\theta} ,\bm{n}'-\bm{n}\rangle - K_{\bullet \bm{n}}(\bm{\theta})\}, 
\end{align*}
where $\bm{n},\bm{n}' \in \Z^d$,
$
K_{0}(\bm{\theta}) = \log \sum_{\bm{j} \in \Z^d} \exp\{\langle \bm{\theta}, \bm{j} \rangle\} 
p_{0}(\bm{j})$ 
and
$
K_{\bullet \bm{n}}(\bm{\theta}) = \log \sum_{\bm{j} \in \Z^d} \exp\{\langle \bm{\theta}, \bm{j} \rangle\} 
p_{\bullet \bm{n}}(\bm{j})$,
at any $\bm{\theta} \in \R^d$ for which the two sums above converge.

We note the exponential tilting requires the existence of the c.g.f.
of the transition probabilities, see Eq. (\ref{e158}) and Eq. (\ref{e159}). This is a light-tail
requirement on the transition distributions. In several problems
of physics the state space is a finite set and so these
c.g.f. do always exist and so there are no restrictions on
the value of the tilting parameter $\theta$. But there are situations
of physics in which the existence of the c.g.f. of the transition
probabilities cannot be guaranteed. This is often the case
when the Markov process refers to quantities such 
as the time-averaged current or activity in e.g. exclusion processes or
to quantities subject to kinetic constraints.

From the above derivations, we can compute the desired expectation
in Eq. (\ref{e129})
with Algorithm \ref{al:tilting} in SM~\cite{SM}, Section \ref{algo}, of 
importance sampling by exponential tilting. We consider 
the practical case with fixed initial initial state.

\subsection{Remarks: optimality and $s$-ensemble}        \label{s30}

This section presents two general remarks related to exponential
tilting. The first one concerns the optimal choice of
the tilting parameter and it is given in Section 
\ref{s35}. The second remark concerns the
closely related technique called $s$-ensemble in the physical
literature, which
is another approach to exponential tilting and which is the subject
of Section \ref{sec:s-ensemble}.

\subsubsection{Optimal tilting parameter under time and space homogeneity}            \label{s35}

The methodology introduced so far does not address the question of the selection the tilting parameter associated to the rare event under consideration. In fact, through the numerical examples in Section~\ref{s4}, we will provide evidence that there is usually an optimal tilting parameter minimizing the sampling error of the numerical estimations. We next show the computation of the optimal tilting parameter for a particular problem as an illustration. In particular, we are now interested in the probability 
of reaching, at some final time $t^\dag$, the interval 
of states
$$\I_{c}(a)=[c-a,c+a] \cap \Z, \; \text{ for some integers } \; 
a \ge 0
\text{ and } c.$$
Consider also that the initial state is fixed $X_0=n_0$, for some
$n_0 \in \Z$ much smaller than $c-a$.
The quantity
of interest is thus
\begin{align}                                       \label{e21}
z_{t^\dag}(\I_{c}(a))= \P[X_{t^\dag} \in \I_{c}(a)].
\end{align}


We further assume that the process is homogeneous in the state space:
the transition probability $p_{\bullet n}$ 
does not depend on $n$ and we denote $p_{\bullet \bullet}=p_{\bullet n}$. This is the random walk
of Section \ref{s33}.
Denote by $K_{\bullet \bullet}$ the c.g.f. of
$p_{\bullet \bullet}$.
We want to determine the exponential tilting
parameter $\theta$ for which $\var_{\P_\theta}
(Z_{t^\dag}(\theta,\I_c(a)))$ is small,
i.e. such that $\E_{\P_\theta}[Z^2_{t^\dag}(\theta,\I_c(a))]$ is small. 
Then Eq. (\ref{e29}) leads to
$$
L_{t^{\dag}}(\theta) = 
\exp\{-\theta (X_{t^{\dag}}-n_0)+ t^{\dag} K_{\bullet \bullet}
    (\theta)\}.
$$
We thus have
\begin{align*}
\E_{\P_\theta}\left[\left\{Z_{t^\dag}(\theta,\I_c(a))\right\}^2\right] & = \E_{\P_\theta}[(I\{ 
    c-a \le X_{t^{\dag}} \le c-a \}
                L_{t^{\dag}}(\theta) )^2] \\
    & \le \E_{\P_\theta}[I\{X_{t^{\dag}} \ge c - a \}
    (\exp\{-\theta (X_{t^{\dag}}-n_0)+ t^{\dag} K_{\bullet \bullet}
    (\theta)\})^2] \\
    & \le (\exp\{-\theta (c-a-n_0) + t^{\dag} K_{\bullet \bullet}
    (\theta)\})^2 \E_{\P_\theta}[I\{X_{t^{\dag}} \ge c-a \} ] \\
    & \le \exp\{-2[\theta (c-a-n_0) - t^{\dag} K_{\bullet \bullet}
    (\theta)]\},
\end{align*}
given that $\theta > 0$ whenever $a>0$.
Strict convexity of $K_{\bullet \bullet}$ implies that
the
above exponent is minimized for
$$t^{\dag} K_{\bullet \bullet}'(\theta)=c- a - n_0 > 0,$$
namely for
\begin{equation}                                \label{e150}  
\E_{\P_\theta}[X_{t^{\dag}} - X_0 \mid X_0 = n_0 ] = c-a- n_0,
\end{equation}
which thus recenters the average of $X_{t^{\dag}}$ towards 
the lower bound of the target interval $\I_c(a)$.


The problem of finding the optimal parameter for arbitrary expectations is not yet solved (e.g. for the case of computing the same estimator  in Eq.~\eqref{e21} but for a process with state-dependent transition probabilities). Nevertheless, expressions like Eq.~\eqref{e150} can be used in an intuitive way to reduce the sampling errors.  For example, in the considered situation
where the target interval $\I_c(a)$ is well above the starting point
$n_0$, any value $\theta>0$ that re-drifts the process 
sufficiently upwards is expected to reduce the Monte Carlo variability. We will elaborate more on this point in Section~\ref{sec:homogeneous_binomial_process}.

\subsubsection{$s$-ensemble}\label{sec:s-ensemble}

This section briefly summarizes the alternative closely related 
importance sampling estimator called
$s$-ensemble. It is an ancillary section that
is not required for the comprehension
of this article.
The $s$-ensemble change-of measure~\cite{hedges2009dynamic,garrahan2009first,jack2010large}
is directly defined at the level of path measures as follows,\footnote{Usually, the $s$-ensemble is defined in a more general manner through processes called ``integrated observables''. In this text, we have chosen to work with a specific case of integrated observable, namely the increment 
$X_t - X_0$, and we refer to~\cite{hedges2009dynamic,garrahan2009first,jack2010large} for a more general definition.}
\begin{equation}\label{eq:s-ensemble_measure}
\P_s[X_0=n_0,\dots,X_{t^\dag}=n_{t^\dag}]=\P[X_0=n_0,\dots,X_{t^\dag}=n_{t^\dag}]\frac{\e^{s\left(n_{t^\dag}-n_0\right)}}{{\cal Z}_{t^\dag}(s)},
\end{equation}
$\forall n_0,\ldots,n_{t^\dag} \in \Z$, where
\begin{equation*}
    {\cal Z}_{t^\dag}(s)=\sum_{n_0,n_{t^\dag}\in \Z} \P[X_0=n_0,X_{t^\dag}=n_t]\e^{s\left(n_{t^\dag}-n_0\right)},
\end{equation*}
at any $s\in \R$ where the sum converges. Thus
${\cal Z}_{t^\dag}$ is the moment generating function\footnote{The specific notation ${\cal Z}$ (instead of $M$ used in other sections) for the moment generating function is typical in the $s$-ensemble literature. So is $s$ (instead of $\theta$) for the tilting parameter.} of
the increment $X_{t^\dag}-X_0$. The parameter $s$ is used to fix the first moment of the increment to some desired 
value $c$ through
\begin{equation}\label{eq:s-ensemble_averages}
    \E_{\P_s}[X_{t^\dag}-X_0]=\frac{d}{d s} \log {\cal Z}_{t^\dag}(s) = c.
\end{equation}

From the definition of the $s$-ensemble probability  measure in Eq.~\eqref{eq:s-ensemble_measure}, we can readily derive the likelihood ratio process of this change-of-measure in the following form,
\begin{equation}\label{eq:s-ensemble_LP}
    L_{t^\dag}(s)=\e^{-s\left(X_{t^\dag}-X_0\right)}{\cal Z}_{t^\dag}(s).
\end{equation}

The $s$-ensemble change-of-measure is similar in form to our exponential tilting, as both techniques bias the original measure
by exponential factors. 
Also, in a process with homogeneous transition probabilities that do not depend on the state of the process, nor on time, both changes of measure are identical (see section~\ref{subsec:exampleI}). Also, a more general definition of exponential tilting, with one tilting parameter per unit of time, can include both the exponential tilt for homogeneous Markov processes of Section \ref{s34} and the $s$-ensemble. However, the $s$-ensemble and exponential tilt as
given in Section \ref{s34}
exhibit relevant differences: the $s$-ensemble draws paths with fixed mean global increment through Eq.~\eqref{eq:s-ensemble_averages}~\cite{claussen2015convex,hartmann2020convex}. Also, the asymptotic properties of the $s$-ensemble make it a useful tool to compute large deviation rates~\cite{hurtado2020building,lecomte2007numerical,perez2019sampling,brewer2018efficient,grassberger2002go,hedges2009dynamic,garrahan2009first,jack2010large,tailleur2009simulation,PhysRevLett.98.195702} and constrained paths in the limit of large times $(t^\dag\to \infty)$~\cite{chetrite2013nonequilibrium,chetrite2015nonequilibrium}. 
On the other side, while transition probabilities in our exponential tilting are obtained through a simple transformation of the transition probabilities of the original process, obtaining the transition probabilities of the $s$-ensemble is not a trivial task; see e.g.~\cite{hurtado2020building,carollo2018making}. Since the focus of this work lies on rare paths within finite time horizons, we do not explore applications of the $s$-ensemble change-of-measure.

\section{Stochastic bridges for Markov processes}\label{sec:backtracking}

The bridge process provides a practical alternative technique to
exponential tilting. It also constructs a new sampling probability measure $\Tilde{\P}$ that makes frequent a given event of 
interest, which is rare under $\P$.
The main idea is to generate a bridge process with fixed 
boundary points or endpoints.
The bridge process is then used for sampling the rare paths that 
possess unlikely pairs of endpoints under the original probability. In fact, as we explain below, the technique can be readily extended to address arbitrary initial and final distributions.
There are various recent applications of this methodology
for sampling rare events in the context of intrinsically out of equilibrium systems~\cite{aguilar2022sampling,majumdar2015effective,wang2022brownian,wang2020exact,de2022optimal,orland2011generating,koehl2022sampling,delorme2016extreme}. The bridge methods are similar in spirit to the transition path sampling algorithms~\cite{van2012dynamical,bolhuis2010trajectory,dellago2009transition}, extensively used in equilibrium molecular dynamics, where paths constrained to both ends are sampled using a Metropolis-Hastings scheme. However, transition path sampling is based on a proposal-rejection scheme like the Metropolis-Hastings algorithm, meaning that generated paths are accepted with some probability. Contrary, all transition paths generated with a bridge process will end in the desired regions by construction.
The generator of the bridge process is obtained conditioning the transition probabilities~\cite{chetrite2013nonequilibrium,chetrite2015nonequilibrium}, which is explained in the following section. As before, we consider processes with discrete state and time spaces. 
With the methods of this section, we always need a fixed
time horizon $t^\dag \ge 1$. We
consider the time $t< t^\dag$, in $\N$,
and the states $n_0,n_{t^\dag},n,n' \in \Z$.

\subsection{Conditioned Markov process}\label{sec:Conditioned_MP}

The bridge process is obtained upon conditioning the original Markov process on passing through particular states at given times. It is possible to sample the stochastic bridges both backward or forward in time, giving rise to two possible generators that we describe below.

\subsubsection{Forward generator}

We can derive transition probabilities that are conditional 
on some fixed final state $X_{t^\dag}=n_{t^\dag}$
through the relation
\begin{align}\label{eq:proof_forward_conditioning_rate}
    \P[X_{t+1}=n'|X_{t}&=n,X_{t^\dag}=n_{t^\dag}] =\frac{\P[X_{t+1}=n',X_{t}=n,X_{t^\dag}=n_{t^\dag}]}{\P[X_{t}=n,X_{t^\dag}=n_{t^\dag}]}  \nonumber \\
    &= \P[X_{t+1}=n'|X_t=n]\frac{\P[X_{t^\dag}=n_{t^\dag}|X_{t+1}=n',X_t=n]}{\P[X_{t^\dag}=n_{t^\dag}|X_{t}=n]} \nonumber \\
    &= \P[X_{t+1}=n'|X_t=n]\frac{\P[X_{t^\dag}=n_{t^\dag}|X_{t+1}=n']}{\P[X_{t^\dag}=n_{t^\dag}|X_{t}=n]}.
\end{align}
We can re-express Eq. (\ref{eq:proof_forward_conditioning_rate}) with
specific notation for transition probabilities 
and change-of-measure kernels
as
\begin{align} \label{eq:proof_forward_conditioning_rate_bis}
    \tilde{p}_{t+1,n}(n'-n;t^\dag,n_{t_\dag}) =
    u_{t+1}(n,n';t^\dag,n_{t^\dag}) \,
    p_{t+1,n}(n'-n),
\end{align}
where
\begin{align}                                           \label{e238}
u_{t+1}(n,n';t^\dag,n_{t^\dag}) =
\frac{\P[X_{t^\dag}=n_{t^\dag}|X_{t+1}=n']}{\P[X_{t^\dag}=n_{t^\dag}|X_{t}=n]}
\end{align}
and
\begin{align*}
 \tilde{p}_{t+1,n}(n'-n;t^\dag,n_{t_\dag}) = 
 \P[X_{t+1}=n'|X_{t}&=n,X_{t^\dag}=n_{t^\dag}].
\end{align*}
Thus, Eq. (\ref{eq:proof_forward_conditioning_rate_bis})
is a special case of Eq.~\eqref{eq:new_transition_prob} (with $q_t=1/u_t$).
By using the transition probabilities on  Eq.~\eqref{eq:proof_forward_conditioning_rate_bis} and by considering
fixed initial state,
we obtain paths that necessarily cross the boundary points $X_0=n_0$ and $X_{t^\dag}=n_{t^\dag}$. We have thus generated a bridge process.
Since the transition probabilities in Eq.~\eqref{eq:proof_forward_conditioning_rate_bis} operate forward in time, we call this procedure the forward generator of the bridge. 

\subsubsection{Backward generator}                                      \label{s412}

Now we construct bridges with transition probabilities that fix the initial state and operate
backward in time. Using manipulations similar 
to those of Eq. (\ref{eq:proof_forward_conditioning_rate}),
we obtain
\begin{align*}
    \P[X_t=n'|X_{t+1}&=n,X_0=n_0] =
     \P[X_{t+1}=n|X_t=n'] \frac{\P[X_{t}=n'|X_{0}=n_0]}{\P[X_{t+1}=n|X_{0}=n_0]},
\end{align*}
namely
\begin{align}\label{eq:proof_backtracking_rate}
    \P[X_t=n'|X_{t+1} =n,X_0=n_0] & = w_{t}(n',n;n_0)
     \P[X_{t+1}=n|X_t=n'] \nonumber \\
     & = w_{t}(n',n;n_0) p_{t+1,n'}(n-n'),
\end{align}
where
\begin{equation}\label{eq:transition_kernel_backtracking}
w_{t}(n',n;n_0)=\frac{\P[X_t=n'|X_0=n_0]}{\P[X_{t+1}=n|X_0=n_0]}
\end{equation}
is the backward change-of-measure kernel. Considering fixed final states ($X_{t^{\dag}}=n_{t^{\dag}}$), the transition probabilities in Eq.~\eqref{eq:proof_backtracking_rate} draw stochastic bridges connecting $X_0=n_0$ and $X_{t^{\dag}}=n_{t^{\dag}}$. The kernels of Eq. \eqref{eq:transition_kernel_backtracking} have different functionality than the change-of-measure 
kernel of Eq. (\ref{e110}). This distinction arises from the backward nature of the generator for the bridge process. 

Both forward Eq.
(\ref{eq:proof_forward_conditioning_rate})
and backward Eq.
(\ref{eq:proof_backtracking_rate})
generators are obtained upon multiplying the original transition probabilities by the change-of-measure kernels $u_{t+1}$ and $w_t$.
These kernels take the form of Doob's $h$-transform, cf. p.
190-195 of ~\cite{asmussen2003applied}. Nevertheless, the probabilities in numerator and denominator of $u_{t+1}$ of
Eq.~\eqref{e238} are efficiently computed through a backward Kolmogorov equation, whereas 
the probabilities 
in numerator and denominator of $w_{t}$
of Eq.~\eqref{eq:transition_kernel_backtracking} are usually computed with a forward Kolmogorov equation. 

Applications that involve sampling bridges with a common initial state (at $t=0$) but multiple final destinations
(at $t=t^\dag$), thus with the Kronecker delta initial 
distrbution $\P[X_0=n]=\delta_{n,n_0}$, are better addressed by the backward generator, as described in \cite{aguilar2022sampling}. The reason is that with the backward generator we need to iterate the forward Kolmogorov equation only once, in order to compute the quantities $\P[X_t=n|X_0=n_0]$, for all relevant values of $n$ and $t$, that are necessary for obtaining
the backward change-of-measure kernel 
Eq. \eqref{eq:transition_kernel_backtracking} and thus for
sampling the bridges. On the other hand, sampling bridges with fixed initial state and multiple final destinations using the forward generator with its change-of-measure kernel in Eq.~\eqref{e238} would require iterating the backward Kolmogorov equation once for each final point of the bridge.
For the same reason, the forward generator is more practical than the backward generator when investigating ensembles of bridges with a fixed final position but varying initial conditions.

Since we focus on problems with fixed initial condition, we will work in the following with the backward generator, also called backtracking method~\cite{aguilar2022sampling}. However, it's worth noting that all the derivations related to the backward generator have their analogous counterparts for the forward generator, so we can make this choice without any loss of generality.

\subsection{Stochastic bridge change-of-measure and likelihood ratio}\label{s42}

Now, we want to use one of the bridge generators of section~\ref{sec:Conditioned_MP} to define the stochastic bridge measure,  an alternative version 
of the generic change-of-measure formulae derived in section~\ref{sec:Backtracking_dt_ds},
which we note as $\tilde{\P}=\P_{n_0}$. To completely define such path measure, we have to specify the statistics of the initial and final conditions. We focus on paths with fixed initial condition, $X_0=n_0$, so that $\P_{n_0}[X_0=n]=\delta_{n,n_0}$. We have considerable freedom in selecting the final distribution, subject to constraints of admissibility associated with absolute continuity (see Section~\ref{subsec:choice_of_final_distribution} for a detailed discussion of this aspect). We call $w_{t^{\dag}}$ the probability function of the final state,
\begin{equation}\label{eq:distribution_last_state_backtrack}
    \P_{n_0}[X_{t^{\dag}}=n]=w_{t^{\dag}}(n;n_0).
\end{equation}
We chose to use the backward generator to draw realizations of the Markov process under $\P_{n_0}$ backward in time, following the transition probabilities specified in Eq.~\eqref{eq:proof_backtracking_rate},
\begin{equation}\label{eq:transition_probabilities_backtracking}
    {\P}_{n_0}[X_t=n'|X_{t+1}=n]=w_{t}(n',n;n_0)\P[X_{t+1}=n'|X_t=n].
\end{equation}

The corresponding likelihood ratio is obtained by multiplication of these
backward
change-of-measure kernels and it is thus given by
\begin{align}\label{eq:likelihood_ratio_backtrack_V0}
    L_{t^{\dag}}(n_0)=\left\{w_0(n_0,X_1;n_0) \, \ldots \, w_{t^\dag-1}(X_{t^\dag-1},X_{t^\dag};n_0)w_{t^\dag}(X_t^\dag;n_0)\right\}^{-1}.
\end{align}
Interestingly, all terms in Eq.~\eqref{eq:likelihood_ratio_backtrack_V0} cancel out excepting those that depend on the final state at time $t^\dag$. 
Therefore, the
expression of the likelihood ratio reduces to
\begin{align}               \label{eq:likelihood_ratio_backtrack}
    L_{t^{\dag}}(n_0)=\frac{h(X_{t^{\dag}};n_0)}{w_{t^{\dag}}(X_{t^{\dag}};n_0)},
\end{align}
where
\begin{equation*}
    h(n;n_0) = \P[X_{t^{\dag}}=n|X_0=n_0].
\end{equation*}
Thus,
for some given function
$g \!: \Z^{t^{\dag}+1} \to \R$,
the importance sampling estimator
of backtracking is given by
\begin{align}                               \label{e948}
Z_{n_0,t^\dag}=g(n_0,X_1,\ldots,X_{t^{\dag}}) 
L_{t^{\dag}}(n_0)
\end{align}
and we have
\begin{align*}                                  
z_{t^\dag} = \E_{{\P}} [ g(n_0,X_1,\ldots,X_{t^{\dag}}) ]  =
\E_{\P_{n_0}} \left[ g(n_0,X_1,\ldots,X_{t^{\dag}}) 
L_{t^{\dag}}  \right] 
= \E_{\P_{n_0}} \left[ Z_{n_0,t^\dag} \right].
\end{align*}

We conclude this section with two remarks.
We first note that with backtracking it is necessary to
fix the time horizon $t^\dag$ in advance and that there
will be only one likelihood ratio random variable,
to be used at all intermediate times $t\in [0,t^\dag]$, 
instead of a complete likelihood ratio process over the time horizon $[0,t^\dag]$.
The second remark concerns the extension of backtracking to the multidimensional state space. It turns out that there
is no conceptual difference when considering a Markov process $\{\bm{X}_t\}_{t \ge 0}$ taking 
individual values in $\Z^d$, for some $d \ge 2$. All formulae of Section 
\ref{s412} and of the present section remain
valid in their given form when the states $\bm{n}_0,\bm{n},\bm{n}'$ represent points of $\Z^d$.

With the above results, we can provide 
Algorithm \ref{al:backtracking} in Section \ref{algo} in SM~\cite{SM} for
importance sampling by backtracking,
for the computation of $z_{t^{\dag}}$.

\subsection{Choice of terminal distribution}\label{subsec:choice_of_final_distribution}

The efficiency of the backtracking method depends on the choice of the final distribution of the new process ($w_{t^{\dag}}$). This distribution has an analogue role to the tilting parameter ($\theta$) of the exponentially tilted measure. 
Let $n,n_0 \in \Z$.
For example, if we choose $w_{t^{\dag}}$ to be equal to the distribution of states at time $t^{\dag}$ with the original process ($w_{t^{\dag}}(n;n_0)=\P[X_{t^{\dag}}=n|X_0=n_0]$), then the new and original measures assign the same weights to paths ($L_{t^{\dag}}=1$), and therefore the change-of-measure will not result in improved efficiency for sampling rare events.

The only restriction concerning the choice of the final distribution
 $w_{t^{\dag}}$ is the absolutely continuity
 $\P\ll \P_{n_0}$.\footnote{Rigorously, it is
 $\P\ll \P_{n_0}$
 on ${\cal F}_{t^\dag} \cap \{X_0=n_0\}$.} 
 This condition is fulfilled if 
\begin{equation}\label{eq:backtracking_absolute_continuity}
      w_{t^{\dag}}(n;n_0)=0\Longrightarrow \P[X_{t^\dag}=n|X_{0}=n_0]=0.
 \end{equation}

As discussed in Section \ref{sec:absolute_continuity}, many applications require that all paths sampled with the new probability measure are also accessible with the original measure (e.g. the equivalence between measures). This stronger constraint is fulfilled when
\begin{equation}\label{eq:backtracking_equivalent}
   w_{t^{\dag}}(n;n_0)=0\Longleftrightarrow \P[X_{t^{\dag}}=n|X_{0}=n_0]=0.
\end{equation}

Choices of the distribution $w_{t^{\dag}}$ fulfilling Eq.~\eqref{eq:backtracking_equivalent} and Eq.~\eqref{eq:backtracking_absolute_continuity} generate \begin{color}{black} unbiased\end{color} changes-of-measure, in the sense that the computed expected values are not affected by systematic errors. On the other side, if there are forbidden states under the new measure that were accessible with the original process, then the errors do not tend to zero as the number of Monte Carlo replications increases. Nevertheless, such errors could be smaller than the sampling errors in cases for which the forbidden areas under the new measure have little relevance for the estimator. This applies to problems that involve transition paths between meta-stable states, where choices such as Kronecker delta distributions ($w_{t^{\dag}}(n;n_0)=\delta_{n,n_{t^{\dag}}}$), violating absolute continuity, can nevertheless be employed for computing estimators with sufficiently small errors, as described in~\cite{aguilar2022sampling}.

\section{Examples and numerical study}\label{s4}

In this section,
numerical comparison between backtracking and exponential tilting are presented through the
following examples: the  binomial process, in Section
\ref{sec:homogeneous_binomial_process}, and a process with state-dependent transition probabilities exhibiting meta-stable states, in Section
\ref{s52}.
In this section we consider times $s<t\le t^\dag$, all in $\N$,
and states $n_0,n_{t^\dag},n,n' \in \Z$.

\subsection{Binomial  Markov process}\label{sec:homogeneous_binomial_process}

Random walks are prototypical toy-models to test methods in non-equilibrium statistical physics. Furthermore,  the extreme statistics of random walks have recently become a subject of intense research due to their wide-ranging applications in finance; see e.g.~\cite{mori2019time,benichou2016joint}. Our work utilizes this random walk example as a basis for applying the derivations discussed in previous sections. Through a simple and analytically calculable example, we can better understand the two Monte Carlo methods.

\begin{figure}
\centering
\includegraphics[scale=0.4]{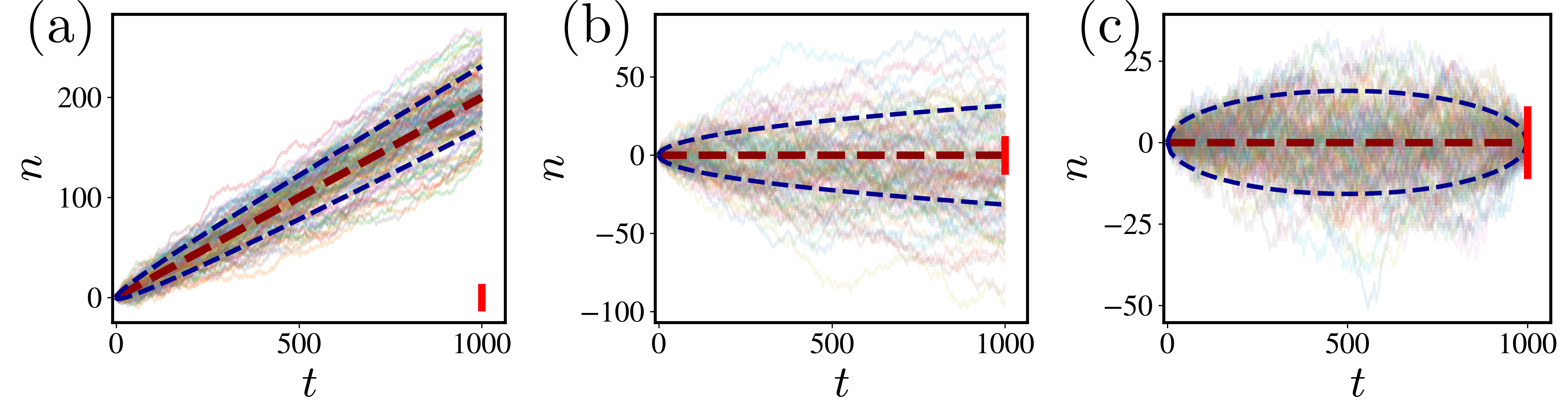}  
\caption{\it In each one of the panels (a), (b) and (c) are shown ensembles of 100 sample paths over the time interval $[0,1000]$. The original binomial process defined through Eq.~\eqref{eq:binomial_UBRW} with $r=0.6$ is drawn in (a), the exponentially tilted process obtained by Eq.~\eqref{eq:transition_prob_exp_tilt_BRW} with $\rho=0$ 
is drawn in (b) and the backtrack processes obtained by Eq.~\eqref{eq:transition_probabilities_backtrack_UBRW} with a Kronecker delta distribution for the last state is drawn in (c). Dashed red lines indicate $\mu_t$ and
dashed blue lines indicate $\mu_t\pm\sigma_t$, where $\mu_t$ denote
expected value and $\sigma_t$ standard deviation, at time $t\in [0,1000]$, both defined in Eq. (\ref{e1000}).
We refer to SM~\cite{SM}, Section \ref{ap:conditioned_moments_binomial}, for details on their calculation. The target interval $\I_0(10)$ from Section~\ref{subsec:exampleI} is shown as a vertical red segment.}\label{fig:trajectories_UBRW}
\end{figure}

\begin{color}{black}The process is defined by the
transition probabilities $p_{\bullet \bullet} = p_{t, n}$,
thus not depending on the state $n \in \Z$, nor on $t \in [0,t^\dag]$, the time. In particular, they are given by \end{color}
\begin{align}        \label{e68}
 p_{\bullet \bullet}(j) & = \begin{cases} 1-r, & \text{if } j =-1, \\
                            r,   & \text{if } j = 1,
 \end{cases}
\end{align}
for some $r\in (0,1)$. The position of the walker follows a binomial distribution (see e.g.~\cite{van2013stochastic}),
\begin{equation*}
    \P[X_t=n|X_s=n']=B\left(\frac{n-n'+t-s}{2},r,t-s\right),
\end{equation*}
where $\forall p \in (0,1)$, $k\in \{1,2,\ldots\}$,
\begin{equation}                                \label{e184}
    B(j,p,k)= \begin{cases} \left( 
    \begin{array}{c} k \\ j 
    \end{array}
    \right)
    p^{j}(1-p)^{k-j}, & \text{if } j=0,1,\ldots,k, \\
                            0,   & \text{otherwise},
 \end{cases}
\end{equation}
are binomial probabilities. Fig.~\ref{fig:trajectories_UBRW}-(a) shows instances of trajectories generated with these transition probabilities, with $r=0.6$ and over the time interval $[0,1000]$ together with the first and second cumulants of the binomial process.
In the following examples, we will consider the problem of estimating the probability of a process departing from the fixed state $n_0\in\Z$ at time $0$ and its subsequent passage through specific 
domains in the time-state space,
within the time interval $[0,1000]$. 
Thus, the probability for the first state is the Kronecker delta given by
$\P[X_0=n] = \delta_{n,n_0}$.

Before presenting the numerical comparisons, in Sections 
 \ref{subsec:exampleI} and \ref{NCII}, let us summarize 
in Sections \ref{s511} and \ref{s512}
how the two importance sampling procedures apply to this particular scenario.

\subsubsection{Likelihood ratio of exponential tilting} \label{s511}

Exponential tilting simplifies substantially when considering
the binomial process.
We find directly the c.g.f.
$$K_{\bullet \bullet} (\theta) = 
\log\left\{r\e^\theta+(1-r)\e^{-\theta}\right\}$$ and thus
the exponential tilting  transition probabilities of Eq.~(\ref{e40}) become 
\begin{equation}\label{eq:transition_prob_exp_tilt_BRW}
 p_{\bullet \bullet}(j,\theta)  = z(n, n+j,\theta) p_{\bullet \bullet}(j) \\
   = \begin{cases} \frac{1-r}{1-r+r \e^{2 \theta}}, & \text{if } j =-1,\\
        \frac{r}{r + (1-r)\e^{-2 \theta}} ,   & \text{if } j = 1.
 \end{cases}
\end{equation}
Thus, the process is stable under exponential tilting,
in the sense that it remains binomial under $\P_\theta$. In consistency with the rest of the text, we consider the fixed initial state $n_0 \in \Z$.
Similar to what is done in Eq.
(\ref{e150}), we obtain the tilting parameter by
 setting the
 conditional expectation of the total run through distance 
 equal to $n_{t^{\dag}}-n_0$, viz. by solving
\begin{equation}                    \label{eq:eq_for_theta_BRW}    
\E_{\P_\theta}[X_{t^{\dag}}- X_0 \mid X_0 = n_0 ] =t^{\dag} \left(
2\frac{r}{r + (1-r)\e^{-2 \theta}}-1\right)=n_{t^{\dag}}-n_0.
\end{equation}
Where the above expectation 
is the one of the binomial distribution.
Note that the capability of fixing the first moment of $X_{t^\dag}-X_0$, which is always true in the $s$-ensemble (cf. Section \ref{sec:s-ensemble} and in
particular Eq. (\ref{eq:s-ensemble_averages})) only holds for the case of exponential tilting because of the simplicity of this particular process (i.e. transition probabilities independent of the state of the process). In Section~\ref{s52}, we will treat another example in which Eq.~\eqref{eq:eq_for_theta_BRW} does not hold.

By defining 
\begin{equation}\label{eq:def_rho}
    \rho=\frac{n_{t^{\dag}}-n_0}{t^{\dag}}
\end{equation} 
and by inverting Eq. (\ref{eq:eq_for_theta_BRW}), we obtain 
\begin{equation}\label{eq:theta_BRW}
    \theta =-\frac{1}{2}\log \left(\frac{r}{1-r}\frac{1-\rho}{1+\rho}\right),
\end{equation}
which is well-defined when $|n_{t^{\dag}}-n_0|<t^{\dag}$
(thus with exclusion of the two monotone sample paths). 

Eq.~\eqref{eq:theta_BRW} implies that it is equivalent to fix the tilting parameter ($\theta$), the average current ($\rho$), or the average final destination ($n_{t^{\dag}}$) of the walker. This map between the tilting parameter and the average current $\rho$ makes it easier to find intuitively values of $\theta$ that will bias paths towards desired regions of the space. Furthermore, this expression makes it explicit the analogy with the canonical ensemble, in which the temperature parameter fixes the energy of the system on average; thus explaining why exponential tilt  methods are also referred to as canonical methods~\cite{chetrite2015nonequilibrium}. Still, the choice of the optimal tilting parameter (or average current) that minimizes the relative error will depend to the specific problem to solve (see e.g. Section~\ref{subsec:exampleI}).

Upon inserting Eq. \eqref{eq:theta_BRW} into Eq. \eqref{eq:transition_prob_exp_tilt_BRW}, we obtain
the exponentially tilted transition probabilities in the form
\begin{equation*}
    p_{\bullet \bullet}(j,\theta)=\begin{cases} \frac{1-\rho}{2}, & \text{if } j =-1,\\
        \frac{1+\rho}{2} ,   & \text{if } j = 1.
 \end{cases}  
\end{equation*}
We note that
these tilted probabilities do not depend on 
$r$. In Fig.~\ref{fig:trajectories_UBRW}-(b), we show instances of trajectories generated with these transition probabilities using $\rho=0$. The homogeneity of the process simplifies the expression of the likelihood 
ratio Eq. (\ref{e29}) to
\begin{align}\label{eq:likelihood_ratio_exponential_tilt_UBRW}
    L_t(\theta) & = \e^{-\theta (X_t-n_0)}
        \prod_{k=1}^t M_{\bullet \bullet} (\theta) 
    =\e^{-\theta (X_t-n_0)}\left\{r \e^{\theta}+(1-r)\e^{-\theta}\right\}^t, \; \text{ for } t=1,\ldots, t^{\dag},
\end{align}
where $M_{\bullet \bullet}$ is the moment generating function of
$p_{\bullet \bullet}$.
We can simplify further Eq.
(\ref{eq:likelihood_ratio_exponential_tilt_UBRW})
by introducing the 
tilting parameter as given in Eq. \eqref{eq:theta_BRW} together with the change-of-variables $q=(1+\rho)/2$ and $N_t=(X_t-n_0+t)/2$, yielding
\begin{equation}\label{eq:likelihood_ratio_exponential_tilt_UBRW_V2}
L_t(q)=\left(\frac{r}{q}\right)^{N_t}\left(\frac{1-r}{1-q}\right)^{t-N_t}, \; \text{ for } t=1,\ldots, t^{\dag},
\end{equation}
where the selected tilting parameter $\theta$ is 
encoded in $q$.

\subsubsection{Likelihood ratio of backtracking} \label{s512}

The binomial distribution of the process simplifies the application of backtracking as well, since the probabilities 
of the backward change-of-measure kernels given in
Eq.~\eqref{eq:transition_kernel_backtracking} are
obtained directly from
\begin{equation}\label{eq:binomial_UBRW}
\P[X_t=n|X_0=n_0]= B\left(\frac{t+n-n_0}{2},r,t\right).
\end{equation}

Therefore, insertion of Eq.~\eqref{eq:binomial_UBRW} into Eq.~\eqref{eq:transition_kernel_backtracking} leads to the transition probabilities under the backtracking measure $\P_{n_0}$ 
through Eq. (\ref{eq:proof_backtracking_rate}) as
\begin{equation}\label{eq:transition_probabilities_backtrack_UBRW}
    \P[X_t = n \mid X_{t+1}=n+j, X_0=n_0]
    =\begin{cases} \frac{1}{2}(1-\frac{n_0-n}{t}), & \text{if } j =-1, \\
    \frac{1}{2}(1+\frac{n_0-n}{t})                        ,   & \text{if } j = 1.
 \end{cases}
\end{equation}
We can freely choose the probability distribution of the states at final time $t^{\dag}$,
i.e. $w_{t^{\dag}}$ defined at Eq.~\eqref{eq:distribution_last_state_backtrack}. When the final distribution is a Kronecker delta, $w_{t^{\dag}}(n;n_0)=\delta_{n,n_{t^{\dag}}}$, the process evolving through Eq.~\eqref{eq:transition_probabilities_backtrack_UBRW} has fixed initial ($n_0$) and final ($n_{t^{\dag}}$) positions. Therefore, since the backtracking can fix the current $\rho$ exactly, cf. Eq.~\eqref{eq:def_rho}, and not in average as the exponential tilting method, it is also referred to as a microcanonical method~\cite{chetrite2015nonequilibrium} (in analogy with the microcanonical ensemble in which the energy is fixed exactly).  In Fig.~\ref{fig:trajectories_UBRW}-(c), we show instances of trajectories generated with these transition probabilities using $w_{t^{\dag}}(n;n_0)=\delta_{n,n_0}$. 
Further, the likelihood ratio of Eq.~\eqref{eq:likelihood_ratio_backtrack} in this case reads
\begin{equation}\label{eq:likelihood_ratio_backtrack_UBRW}
    L_{t^{\dag}}(n_0)=\frac{B\left(\frac{t^{\dag}+X_{t^{\dag}}-n_0}{2},r,t^{\dag}\right)}{w_{t^{\dag}}(X_{t^{\dag}};n_0)}.
\end{equation}

\subsubsection{Numerical comparisons I: homogeneity and local time target}\label{subsec:exampleI}

Our first numerical experiment compares
exponential tilting and backtracking for obtaining
the probability that the process departing from state 
$n_0=0$ hits the target interval $\I_0(10)=[-10,10]$ at terminal time $t^\dag=1000$
with a positive bias obtained by setting $r=0.6$. 
This is a simple example that can be solved analytically and we indeed obtain
\begin{equation}\label{eq:first_moment}
    z_{t^{\dag}}(\I_0(10))=\P[X_{t^{\dag}}\in \I_0(10)]=\sum_{n=-10}^{10} \P[X_{t^{\dag}}=n|X_0=0]=7.543\times10^{-10}.
\end{equation}
The objective is thus to benchmark the two Monte Carlo methods when the quantity of interest is available. Moreover, this example enables us to assess the importance of appropriate selecting the tilting parameter and the terminal distribution,
for exponential tilting and backtracking respectively.
We illustrate that their careful selections are 
crucial for the efficiency of these methods.

The homogeneity of the process and the simplicity of the target make that we can actually avoid the generation of the full
trajectory and only generate, by importance sampling, the last state of the process ($X_{t^{\dag}}$). We can then define the importance sampling estimators of exponential 
tilting and backtracking respectively as
\begin{align}                                       \label{e55}
Z_{\theta, t^\dag} (\I_0(10)) & = I_{\I_0(10)}(X_{t^{\dag}} ) L_{t^{\dag}}(\theta)\;\text{ and } \quad  Z_{n_0, t^\dag} (\I_0(10))  = I_{\I_0(10)}(X_{t^{\dag}} ) L_{t^{\dag}}(n_0).
\end{align}
Then, the probability of ending in $\I_0(10)$ is
the expectation of these two Monte Carlo
estimators w.r.t. their importance sampling distributions, precisely 
\begin{align}                                   \label{e89}
z=z_{t^\dag}(\I_0(10))  =\E_{\P_{\theta}}[Z_{\theta, t^\dag} (\I_0(10))]= \E_{\P_{n_0}}[Z_{n_0, t^\dag} (\I_0(10))].
\end{align}

\begin{figure}
\centering
 \includegraphics[scale=0.45]{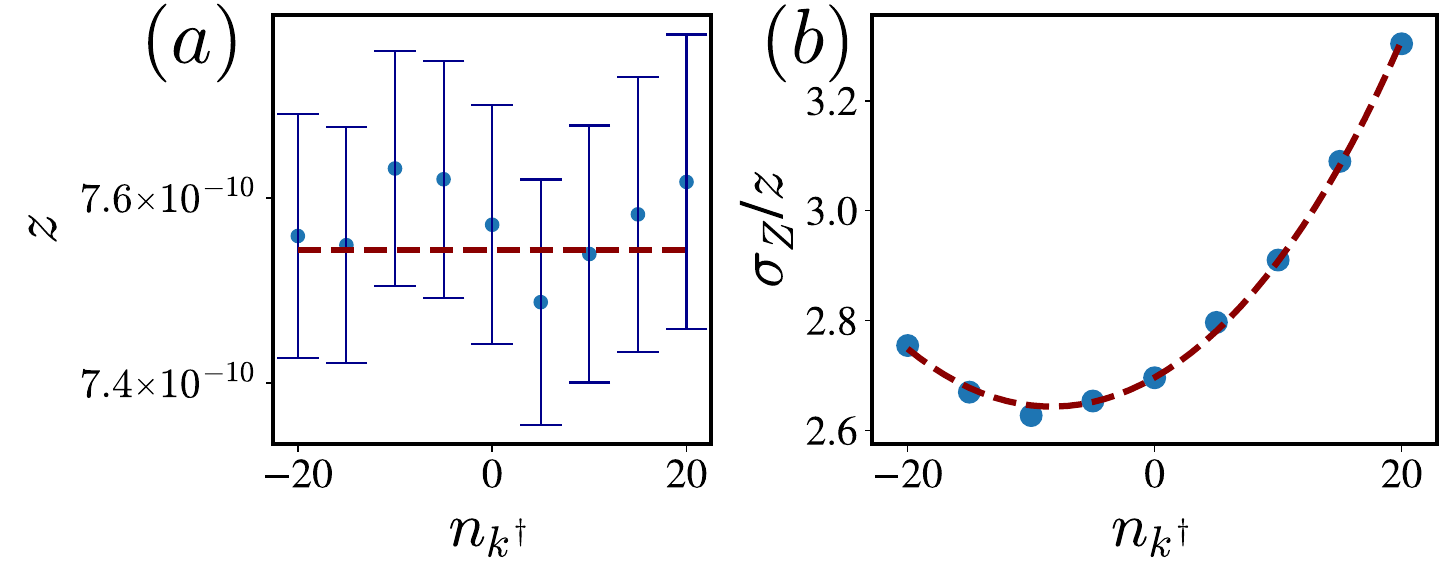}
    \caption{\it Panel (a) shows the exponential tilt estimators 
    of $z_{t^{\dag}}(\I_0(10))$ based on $10^5$ realizations and for different values of the average arrival state ($n_{t^\dag}$). 
    Fixing the arrival state on average is equivalent to fixing of the tilting parameter $\theta$ in Eq.~\eqref{eq:theta_BRW}, $\rho$ in Eq.~\eqref{eq:def_rho} and $q$. The dashed line shows the exact value and the error bars shows the $95\%$ asymptotic normal confidence intervals. Panel (b) shows the Monte Carlo relative error in blue points together with the theoretical values in dashed red line. The minimal relative error is obtained by fixing the mean arrival state equal to the extreme of the target interval ($n_{t^{\dag}}=-10$). }\label{fig:example1_exponential_tilt}
\end{figure}

{\bf For exponential tilting}, we generate random arrival states from the binomial distribution and then compute their mean weighted by the likelihood ratios in Eq.~\eqref{eq:likelihood_ratio_exponential_tilt_UBRW_V2}. 
Algorithm \ref{al40} in SM~\cite{SM}, Section \ref{algo}, generally given for
$\I_c(a)$, $a \ge 0$ and $c$ integers,
follows directly from Eq.
(\ref{e55}) and Eq. (\ref{e89}).
In Fig. \ref{fig:example1_exponential_tilt}-(a) we show that, 
for various values of the tilting parameter,
the importance sampling estimator with the exponential tilted measure is in close agreement with the true analytical value. 

We can also compute the second moment of the exponential tilt estimator as
\begin{equation}                                \label{e117}
    z_2=\E_{\P_\theta}\left[I_{\I_0(10)}(X_{t^{\dag}})\{L_{t^{\dag}}(\theta)\}^2\right]=\sum_{n=n_0-t^{\dag}}^{n_0+t^{\dag}} I_{\I_0(10)}(n)\{L_{t^{\dag}}(\theta)\}^2 B\left(\frac{t^{\dag}+n-n_0}{2},r,t^{\dag}\right).
\end{equation}
Eq. (\ref{e117}) together with Eq.~\eqref{eq:first_moment} allow
us to compute the relative error $\sigma_Z/z$, where
$\sigma_Z=\sqrt{z_2-z^2}$.
This relative error is proportional to the square of the number of Monte Carlo replications required for target precision in the calculations, cf. p. 158-159~\cite{asmussen2007stochastic}. Minimizing the relative error enhances efficiency, in the sense as fewer realizations are necessary 
in order to reach a desired level of precision. 

It is shown in Fig. \ref{fig:example1_exponential_tilt}-(b) that empirical and theoretical values for the relative error are in agreement. Furthermore, Fig. \ref{fig:example1_exponential_tilt}-(b) shows that the minimal relative error is obtained by the tilting parameter that places the average final position at the lower bound of the target interval ($n_{t^\dag}=-10$), this in agreement with the derivations of Section~\ref{s35}.

{\bf For backtracking}, we need to choose the distribution of the last states ($w_{t^{\dag}}$). As discussed above for the exponential tilt, it is simpler to consider the distribution of the number of positive jumps by the final time ($N_{t^\dag}$), which is given by 
\begin{equation*}
    f_{t^{\dag}}(n)=\P[N_{t^\dag}=n].
\end{equation*}
This distribution is related to the one of the terminal states through
\begin{equation*}
    f_{t^{\dag}}(n)=w_{t^{\dag}}(2n+n_0-t^{\dag};n_0).
\end{equation*}
We consider a one parameter family of uniform
distributions, 
\begin{equation}\label{eq:distribution_final_positions_backtracking}
    f_{t^{\dag}}(n)=\begin{cases} \frac{1}{D+1}, & \text{if } n \in[\frac{t^{\dag}}{2}-D,\frac{t^{\dag}}{2}+D],\\
    0, & \text{otherwise}.
 \end{cases}
\end{equation}
The support of the distribution
Eq. \eqref{eq:distribution_final_positions_backtracking} 
can be trivially modified and so we can easily 
examine scenarios where the new sampling measure does not adhere to the absolute continuity requirement. In particular, selecting $D<t^{\dag}$ violates $\P\ll\P_{n_0}$ and leads to the addition of bias errors (refer to Sections \ref{sec:absolute_continuity} and ~\ref{subsec:choice_of_final_distribution} for details). 

\begin{figure}
\centering
 \includegraphics[scale=0.45]{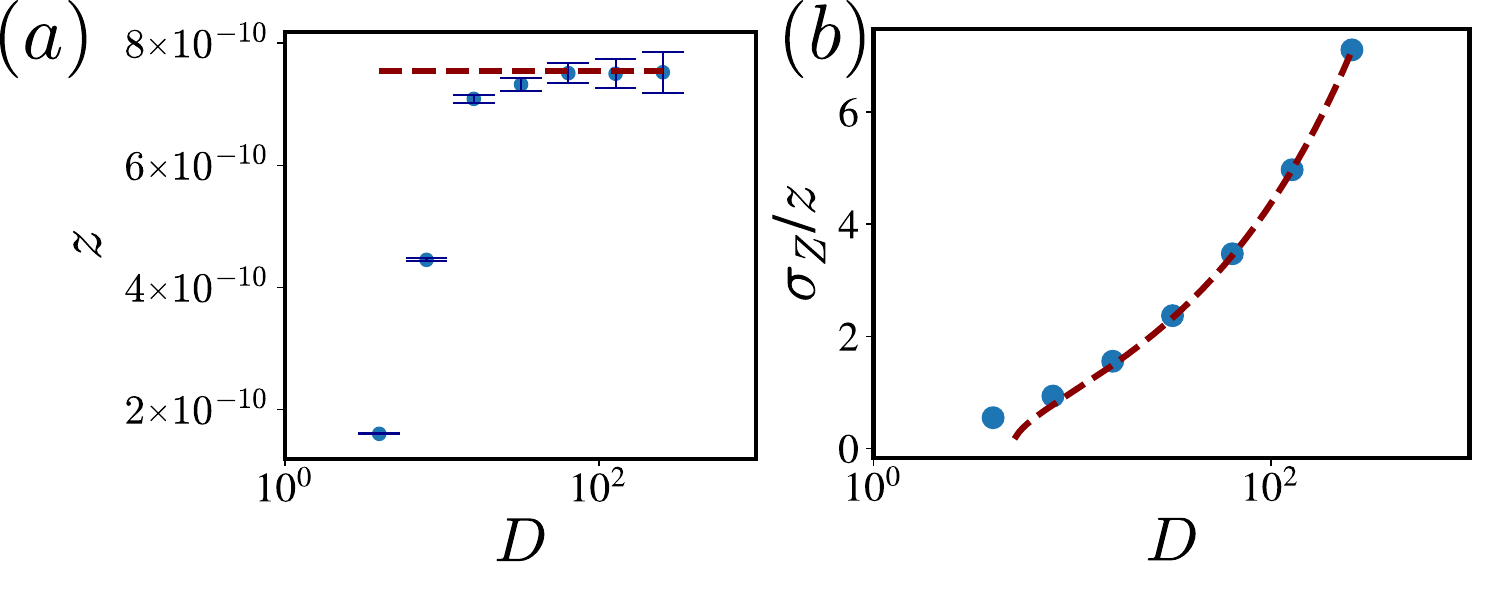}
    \caption{\it Panel (a) shows the results of the estimation of $z=\P[X_{t^{\dag}}\in \I_0(10)]$ by using the backtracking measure, with $10^5$ generations and with different radii $D$ of the support of the uniform distribution of the final state, given in Eq.~\eqref{eq:distribution_final_positions_backtracking}. The red dashed line indicates the exact value and the error bars provide
    $95\%$ asymptotic normal confidence intervals. Panel (b)
    shows the Monte Carlo relative error, with blue dots, together with the theoretical value, with the dashed red line. }\label{fig:example1_backtracking}
\end{figure}

Algorithm \ref{al50} in SM~\cite{SM},  provides the detailed implementation of backtracking for the general target interval
$\I_c(a)$, with any integers $a \ge 0$ and $c$.
The results of the Monte Carlo study in
Fig. \ref{fig:example1_backtracking}-(a) show that importance sampling by backtracking is in substantial agreement with the analytical value, for terminal distributions that include the target region, viz. for $D>10$. But when $D<10$, the backtracking measure forbids states that are important for the estimation of $z$. Such 
values of $D$ lead to \begin{color}{black} bias (systematic)\end{color} errors. In contrast with this, 
when $D\ge10$, forbidden areas do not affect the backtracking estimator, even without fulfillment of absolute continuity.

Then,
Fig. \ref{fig:example1_backtracking}-(b) shows the influence of the width of the support of the terminal distribution ($2D$) on the relative error. As before, simulations are compared with theoretical values, obtained with the formula
\begin{align*}
    z_2& =\E_{\P_{n_0}}\left[I_{\I_0(10)}(X_{t^{\dag}})\{L_{t^{\dag},n_0}(X_{t^{\dag}})\}^2\right]  
    = 2D \sum_{n=n_0-t^{\dag}}^{n_0+t^{\dag}} I_{\I_0(10)}(n) \left\{B\left(\frac{t^{\dag}+n-n_0}{2},r,t^{\dag}\right)\right\}^2.
\end{align*}
We see that the sampling errors are proportional to $D$. However, even if for $D<10$ we observe the smaller statistical errors, discrepancies with the analytical value are bigger due to systematic errors (see  Fig.~\ref{fig:example1_backtracking}-(a)).

\subsubsection{Numerical comparisons II: homogeneity and extended time target}  \label{NCII}

For the second numerical example, we evaluate the probability that the binomial process hits a target extended in time: 
\begin{equation}\label{eq:extended_time_target}
    z_{t_1,t_2}(\I_0(10))=\P[\exists s\in[t_1,t_2]: X_s\in \I_0(10)],
\end{equation}
where $0 \le t_1 < t_2 \le t^\dag$.
As in the previous example, the process is biased towards the positive direction with $r=0.6$. In this case, the hitting probability is not trivial to obtain analytically. Thus,  numerical techniques are the preferred option to tackle the problem. Moreover, since the target can be hit at different times, we need to simulate the process at intermediate times.

{\bf For exponential tilting}, we follow Algorithm~\ref{al:tilting} in SM~\cite{SM}, Section \ref{algo}, where we choose the tilting parameter according Eq.~\eqref{eq:def_rho} and
    Eq. \eqref{eq:theta_BRW} such that the tilting parameter sets the mean final state of trajectories equal to the center of the interval $\I_c(a)$, namely $c$. Also, we can compute the estimator as 
    \begin{align}\label{eq:estimator_exponential_tilted_example2}
        \hat{z}_{t_1,t_2,m}(\I_c(a)) 
        & =\frac{1}{m}\sum_{i=1}^m I\left\{\sum _{s=t_1}^{t_2}I_{\I_c(a)}\left(X^{(i)}_{s}\right)>0\right\}L_{t^{\dag}}(X_{t^{\dag}},
        \theta) \nonumber \\ 
        & =\frac{1}{m}\sum_{i=1}^m I\left\{\sum _{s=t_1}^{t_2}I_{\I_c(a)}\left(X^{(i)}_{s}\right)>0\right\}\left(\frac{r}{q}\right)^{N^{(i)}_{t^{\dag}}}\left(\frac{1-r}{1-q}\right)^{t^{\dag}-N^{(i)}_{t^{\dag}}},
    \end{align}
where 
    \begin{equation*}
        N_{t^{\dag}}^{(j)}=\frac{t^{\dag}+X^{(j)}_{t^{\dag}}}{2}.
    \end{equation*}

{\bf For backtracking}, we use Algorithm~\ref{al:backtracking}
in SM~\cite{SM}, Section \ref{algo}, with the following uniform terminal distribution,
\begin{equation*}
    w_{t^\dag}(n;n_0)=\begin{cases} \frac{1}{D}, & \text{if } n \in \I_c(a),\\
    0, & \text{otherwise},
    \end{cases}
\end{equation*}
with $D=2a$.
Also, we can compute the Monte Carlo estimator of 
    of $z_{t_1,t_2}(\I_c(a))$, given in Eq.~\eqref{eq:extended_time_target}, by
    \begin{align*}
        \hat{z}_{t_1,t_2,m}(\I_c(a)) &=\frac{1}{m}\sum_{i=1}^m  I\left\{\sum _{s=t_1}^{t_2}I_{\I_c(a)}\left(X^{(i)}_{s}\right)>0\right\} L_{t^{\dag},n_0}\left(X^{(i)}_{t^{\dag}}\right) \nonumber \\ &=\frac{2D}{m}\sum_{i=1}^m  I\left\{\sum _{s=t_1}^{t_2}I_{\I_c(a)}\left(X^{(i)}_{s}\right)>0\right\}B\left(N_{t^{\dag}}^{(i)},r,t^{\dag}\right),
        \end{align*}
        where $B$ is the binomial probability function of Eq. (\ref{e184}).

\begin{figure}
    \centering
    \includegraphics[scale=0.45]{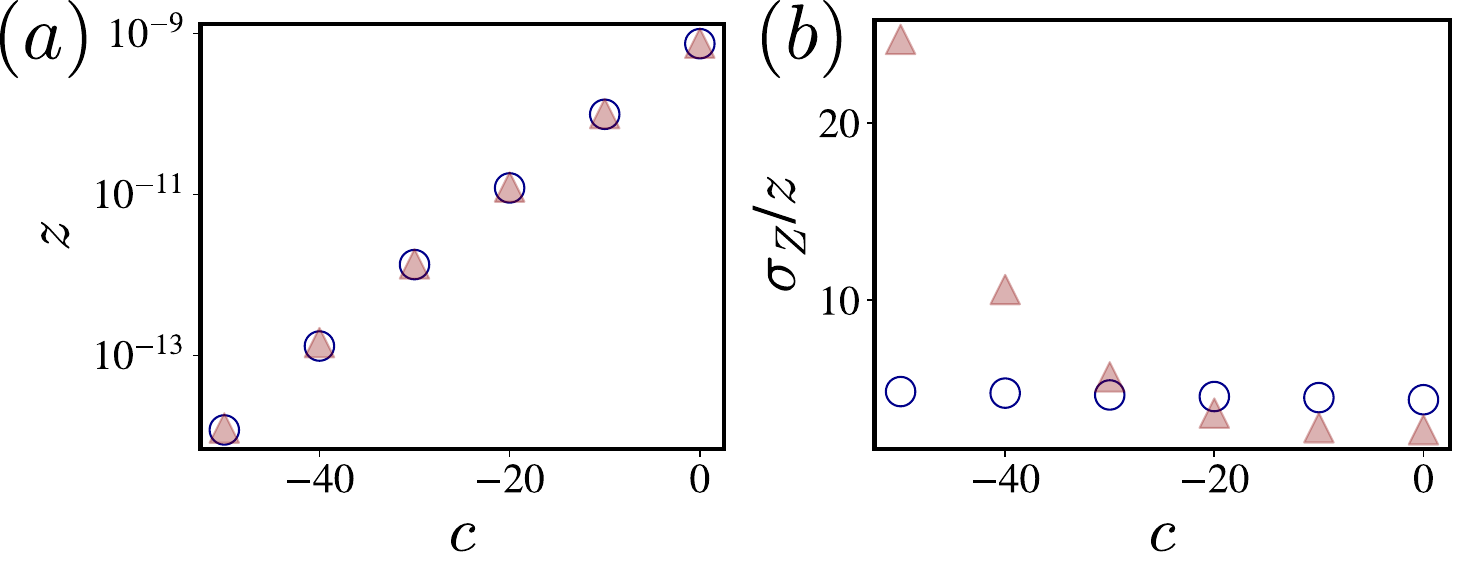}
    \caption{\it Panel (a) shows the results for the estimation of $z=z_{t_1,t_2}(\I_0(10))$ of
    Eq.~\eqref{eq:extended_time_target} by backtracking in blue circles and by exponential tilting in red triangles, always with $10^5$ generations and with varying values of the center of the target interval $c$. Panel (b)
    shows the Monte Carlo relative error obtained by these two methods. Both methods provide good estimations of $z$, with similar small Monte Carlo errors. Both methods display bounded relative error, 
    which however increases with exponential tilting
    for small values of the center of the target $c$, viz. as
    the target becomes more unlikely.}
\label{fig:target_extended_in_time}
\end{figure}

In Fig.~\ref{fig:target_extended_in_time}-(a) we show that the sample average for different values of the center $c$ computed with the two methods agree within errors. In Fig.~\ref{fig:target_extended_in_time}-(b) we also show that the relative errors in both methods have the same order of magnitude. However, the relative error for the exponential tilt increases as the event becomes rarer whereas the errors in the backtracking method are more constant. This result is surprising provided that the backtracking process is biased (because our particular choice of $w_{t^\dag}$ in Eq.~\eqref{eq:likelihood_ratio_backtrack} does not fulfill absolute continuity) whereas the exponential tilted method is unbiased.

\subsection{Process with meta-stable states}     \label{s52}

In this section we study a more sophisticated and practical Markovian process where the
transition probabilities in Eq.
(\ref{e68}) do depend on the value of the state $n$.
Precisely, they take the logistic form
\begin{align}\label{eq:PT_metastable}
 p_{\bullet n}(j) & = \begin{cases} \left\{1+\e^{j\nu{n(n-\ell)(n+\ell)}}\right\}^{-1} , & \text{for } j =-1,1, \\
                            0,   & \text{otherwise},
 \end{cases}
 \end{align}
where $n$ and $\ell>0$ are integers and $\nu>0$ is real. 
We note that transition probabilities in Eq.~\eqref{eq:PT_metastable} define 
a process over a binomial tree (i.e. with stepwise upwards or downwards unit changes), 
since $p_{\bullet n}(1) + p_{\bullet n}(-1) = 1$. Additionally, this Markov 
process exhibits two meta-stable states at positions $n=\pm \ell$. This means that trajectories tend to oscillate around these two states and it is exceptional to observe a transition from one of these states to the other one. The parameter $\nu$ tunes the robustness of the meta-stable states. If $\nu=0$, then the process is an unbiased random walk without meta-stable states. At the opposite, namely at the limit $\nu\to\infty$, trajectories evolve by forming a straight line (deterministically) towards either $\ell$ or $-\ell$. 

\begin{figure}
    \centering
    \includegraphics[scale=0.47]{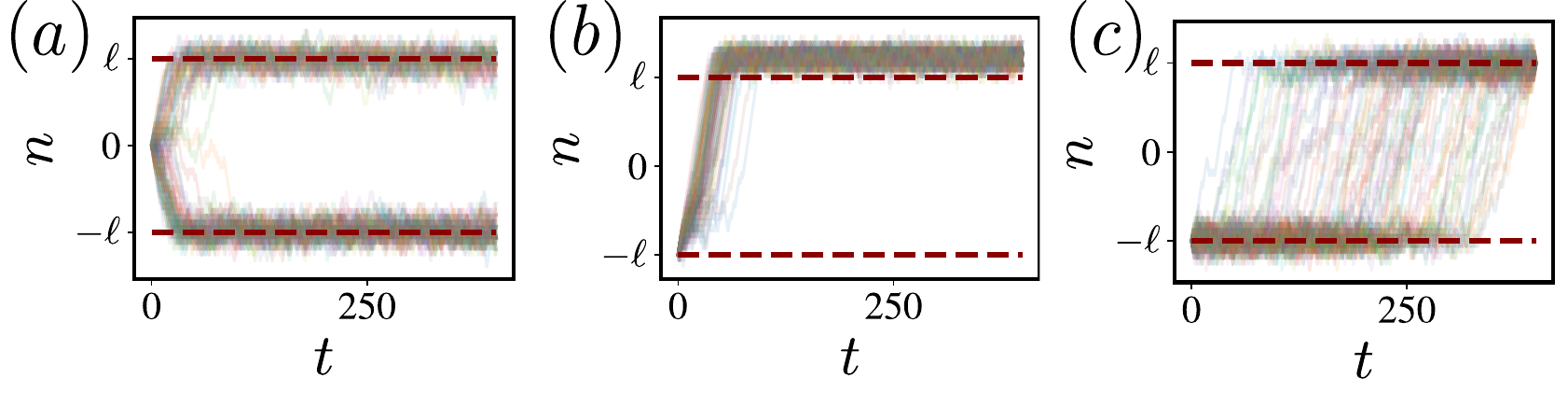}
    \caption{\it Panels (a), (b) and (c) show ensembles of 100 sample paths for $t\in \{0,\dots,400\}$ and parameters $\nu=10^{-3}$ and $\ell=15$. Trajectories in (a) are generated with the original process with transition probabilities depending on  the states through Eq.~\eqref{eq:PT_metastable}. All paths depart from position $n=0$ $\left(\P[X_0=n]=\delta_{0,n}\right)$. The meta-stable states $n=\pm \ell$ are indicated by the two horizontal dashed red lines. Due to the symmetry of the transition rates around $n=0$, on average, one half of the trajectories evolve towards the meta-stable state $n=\ell$ and the other half evolve towards the state $n=-\ell$. No transitions between the meta-stable states were observed within the given time horizon. Trajectories in (b) are generated under the tilted measure
    of Eq.~\eqref{eq:PT_exp_tilt} with tilting parameter $\theta=1$.  We note that positive values of the tilting parameter favor transitions towards the positive meta-stable state. We note that the location of the meta-stable state of the tilted process is slightly above the one of the original process. Trajectories in (c) are obtained by backtracking and transitions over a wide range of intermediate times can be observed. 
    It is shown that the backtracking process does respect the position of the meta-stable states of the original process.}
\label{fig:trajectories_metastable}
\end{figure}

 Furthermore, the roles of $\ell$ and $\nu$ can be understood intuitively from a physical perspective. The transition probabilities in Eq.~\eqref{eq:PT_metastable} can be seen as a discretization of the trajectory of a particle that, at position $n$, is
subject to the force 
$$-\tanh \frac{\nu{n(n-\ell)(n+\ell)}}{2},$$ in the overdamped limit i.e. in a high viscosity regime where inertia can be neglected. In Fig.~\ref{fig:trajectories_metastable}-(a) we show typical trajectories for this process.

In order to statistically characterize these transitions, we make use of our two 
change-of-measure strategies.

{\bf For exponential tilting}, 
by using Eq.~\eqref{e40} we obtain 
the following transition probabilities for any fixed value of the 
tilting parameter $\theta \in \R$,
\begin{align}\label{eq:PT_exp_tilt}
 p_{\bullet n}(j,\theta) & = \begin{cases} \left\{1+\e^{j\left[\nu{n(n-\ell)(n+\ell)}-2\theta\right]}\right\}^{-1} , & \text{for } j =-1,1, \\
                            0,   & \text{otherwise}.
 \end{cases}
 \end{align}
The parameter $\theta$ breaks the symmetry of the transition rates around $n=0$ and effectively biases paths either towards the meta-stable state in the positive position $\ell$ (for $\theta>0$) or in the negative position $-\ell$ (for $\theta<0$). 
Moreover, Eq. (\ref{eq:PT_exp_tilt}) tells that
the precise positions of the meta-stable states are not preserved and depend on the value of $\theta$. This fact is illustrated numerically in 
Fig.~\ref{fig:trajectories_metastable}-(b).

In this case, the likelihood ratio process of the exponentially 
tilted measure depends on the whole 
Markov process and it is obtained by Eq.~\eqref{e29}.

{\bf For backtracking}, we cannot use an analytical form for the transition probabilities and we have to evaluate 
the backward change-of-measure kernel of
Eq.~\eqref{eq:transition_kernel_backtracking} numerically. In principle, this evaluation would require to compute the probabilities $\P[X_t=n|X_0=n_0]$,
for all values of $n$ and $t$, through the iteration of the forward Kolmogorov equation 
\begin{equation}\label{eq:forward_KE}
    \P[X_{t+1}=n|X_0=n_0]= p_{\bullet n+1}(-1) \P[X_t=n+1|X_0=n_0]+p_{\bullet n-1}(1) \P[X_t=n-1|X_0=n_0].
\end{equation}
In order to ensure numerical tractability of Eq. \eqref{eq:forward_KE}, boundary conditions are introduced to the system. We set $\ell_\text{max}$ and $-\ell_\text{max}$ as boundaries such that 
$$p_{\bullet -\ell_\text{max}}(-1)=p_{\bullet \ell_\text{max}}(1)=0 \text{ and } p_{\bullet -\ell_\text{max}}(0)=p_{\bullet \ell_\text{max}}(0)=\left\{1+\e^{-\nu{n(n-\ell_\text{max})(n+\ell_\text{max})}}\right\}^{-1}.$$ 
These conditions imply that particles are unable to cross the boundaries, but they can avoid
jumping, viz. remaining in the same state, precisely at the boundaries. This approximation does not introduce significant systematic errors in the solution as long as the probability for the process to reach the boundaries is negligible 
(viz. $\P[X_t=\pm\ell_\text{max}] \simeq 0$, 
for $t=1,\ldots,t^\dag$). 
Although it is possible to introduce alternative approximations that avoid the use of Eq. \eqref{eq:forward_KE}, they are not employed in this work. For more details on these alternative approaches, we refer the interested reader to ~\cite{aguilar2022sampling}. In Fig.~\ref{fig:trajectories_metastable}-(c) it is shown that the process generated by backtracking can be used to sample paths connecting meta-stable states.

\subsubsection{Numerical comparisons III: inhomogeneity and local time target}\label{subsec:exampleIII}

 As with the homogeneous binomial process
 in Section~\ref{subsec:exampleI}, we present another example for which the desired quantity can be evaluated analytically.
 Let $\ell$ be a positive integer.
 We are interested in the probability that a path starting from one of the two meta-stable states, precisely $-\ell$, hits a one-sided interval that includes the other meta-stable state, $\ell$, at some terminal time $t^{\dag}$. This is the probability
$$z(A)=\P[X_{t^{\dag}}\in A|X_0=-\ell],$$
where $A=[a,\infty)$ and the integer $a$ is such that $-\ell\notin A$ and $\ell \in A$,
i.e. such that $-\ell < a < \ell$. In fact,
this problem can be solved by the direct evaluation of
$\P[X_{t^{\dag}}=n|X_0=-\ell]$ at all integers $n\in A$, by means of Eq.~\eqref{eq:forward_KE}, 
and then by summing, because
\begin{equation}\label{eq:computation_of_z}
    z(A)=\sum_{n\in A} \P[X_{t^{\dag}}=n|X_0=-\ell].
\end{equation}
In this study,
we first estimate $z(A)$ through an ensemble of paths generated by 
backtracking, as described in Algorithm~\ref{al:backtracking}
in SM~\cite{SM}, Section \ref{algo}.

The terminal distribution is the uniform one, given by
\begin{equation}                                        \label{e583}
    w_{t^\dag}(n;-\ell)=\begin{cases} \frac{1}{2D+1}, & \text{if } n \in \I_\ell(D),\\
    0, & \text{otherwise},
    \end{cases}
 \end{equation}
 for some positive integer $D$.

Our first experiment considers extensions of the temporal horizon of the simulation by backtracking. We thus generate sample paths by backtracking at times $t=0,\dots,t^*$, this
for several values of $t^*\ge t^\dag$. The aim is to determine the effect of the terminal time
on the accuracy of the Monte Carlo estimator of $z(A)$. So the terminal distribution
in Eq. (\ref{e583}) is considered at time $t^*$, instead of $t^\dag$.

\begin{figure}
    \centering
    \includegraphics[scale=0.45]{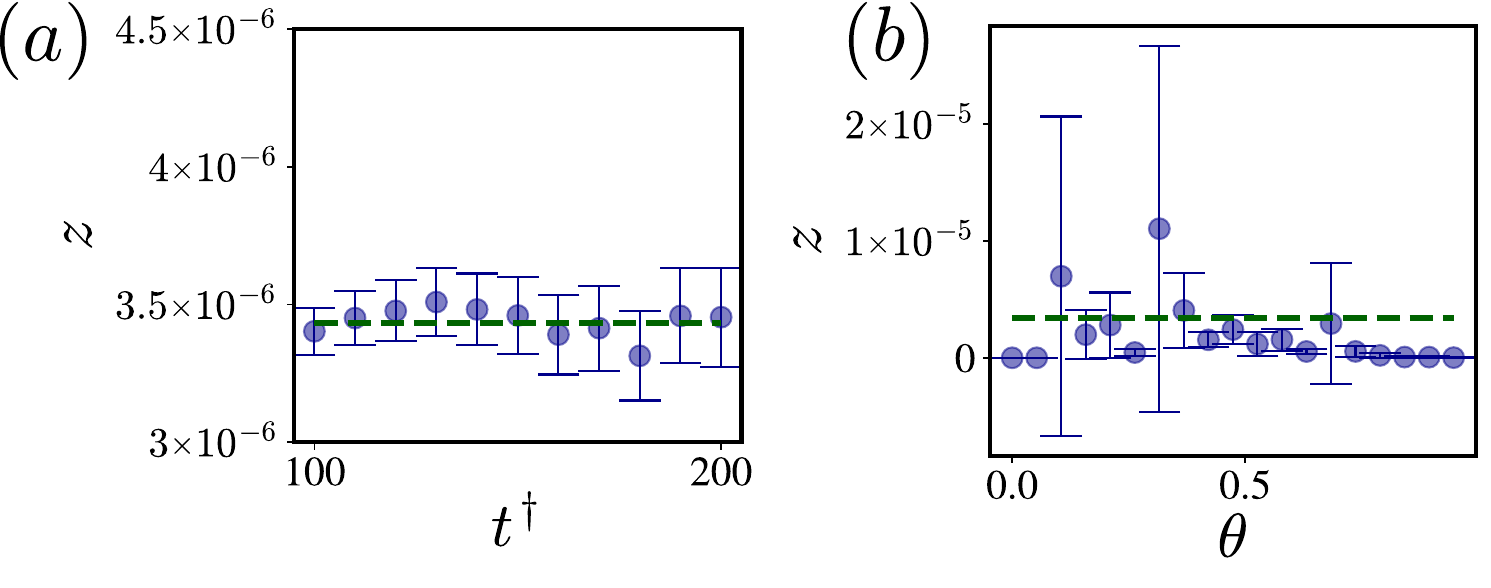}
    \caption{\it Estimations of $z(A)=\P[X_{t^{\dag}}\in A|X_0=-\ell]$, for $A=[\ell-2,\infty]$, $\ell=15$, $\nu=10^{-3}$, $t^{\dag}=100$, by backtracking in panel (a) and by exponential tilting in panel (b), always with $10^4$ generations. Error bars show the $95\%$ asymptotic normal confidence intervals.}
    \label{fig:exp_3}
\end{figure}

In Fig.~\ref{fig:exp_3}-(a) we see that the Monte Carlo estimations of $z(A)$ 
are close to the value computed analytically 
by Eq.~\eqref{eq:computation_of_z}. We also see that the Monte Carlo 
errors increase as $t^*$ increases away from $t^{\dag}$.

\begin{figure}
    \centering
    \includegraphics[scale=0.45]{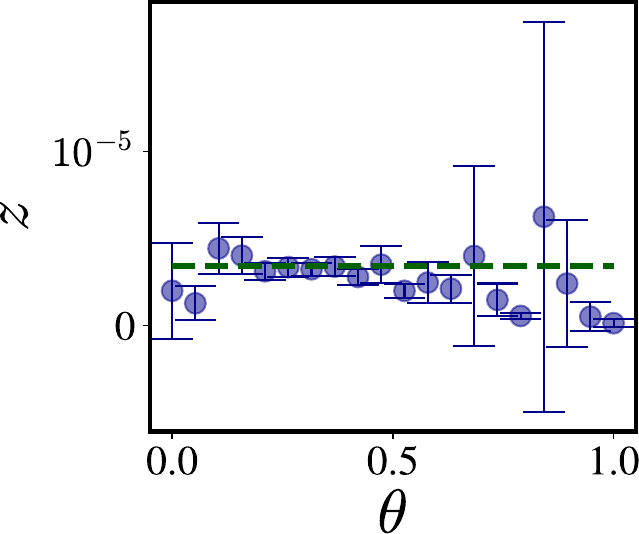}
    \caption{\it Estimations of $z(A)=\P[X_{t^{\dag}}\in A|X_0=-\ell]$, for $A=[\ell-2,\infty]$, $\ell=15$, $\nu=10^{-3}$, $t^{\dag}=100$ and by exponential tilting with $10^6$ generations. Error bars show the
    $95\%$ asymptotic normal confidence intervals.}
    \label{fig:exp_3_more_realiz}
\end{figure}

In Fig.~\ref{fig:exp_3}-(b) we show the computation of $z(A)$ over an ensemble of paths generated by Algorithm \ref{al:tilting} in SM~\cite{SM}, Section \ref{algo}, of exponential tilting, 
for different values of the tilting parameter $\theta$. 
In contrast with backtracking, the exponential tilting estimator does not seem to converge 
to  $z(A)$. Furthermore, certain 
sample paths
possess large fluctuations, visible in the error bars of Fig.~\ref{fig:exp_3}-(b). Thus,
we observe that the distribution of the exponential tilting estimator exhibits heavy tails. Consequently, a reliable estimation for the first moment necessitates a 
very large number of simulations. As this number increases, we observe a notable improvement of the exponential tilting estimation in Fig.~\ref{fig:exp_3_more_realiz}, in which
exponential tilting is based on a larger number of simulations than in Fig.~\ref{fig:exp_3}-(b). Fig.~\ref{fig:exp_3_more_realiz} tells that the optimal tilting parameter $\theta$ is located around $0.3$. 
Large discrepancies between the numerical estimator and the analytical result are observed for values of $\theta$ close to one, pointing out the importance of using the optimal tilting parameter for 
this type of application.

The example presented in this last section offers novel insights into the suitability of backtracking for addressing problems involving transition paths between meta-stable states. Remarkably, 
backtracking, although biased due to the violation of absolute continuity of measures, proves to be more effective in such scenarios than exponential tilting, which is unbiased.
\section{Final remarks}               \label{s5}

 Problems that require sampling rare trajectories within a feasible amount of time appear in various scientific disciplines (such as physics, mathematical biology and finance). In this article, we analyze two paradigms of methods for sampling rare trajectories
 of Markov processes: exponential tilting and stochastic bridge. Our main contribution is to show that these two methods can be re-expressed within the general theory of change-of-measure and importance sampling.
 Through practical numerical examples,
 we illustrate the applicability of both Monte Carlo methods to the computation of
 probabilities of rare trajectories.
Moreover, we provide various Monte Carlo algorithms in detailed form, so to make them directly 
accessible applied scientists.

We have demonstrated, both theoretically and through simulations, that techniques relying on sampling stochastic bridges may lead to systematic errors in the estimation of averages (even when the bridge generator itself is unbiased) if absolute continuity is not satisfied. To the best of the authors' knowledge, this represents a novel finding to consider when employing such methods. It highlights that calculations with a small Monte Carlo error could prove significantly inaccurate in the absence of absolute continuity.

For the case of the binomial process, we provide numerical 
 evidence that the relative errors of these two sampling 
 strategies have small and comparable order. In problems involving transition paths between meta-stable states, stochastic bridges appear to assign consistent values to the estimators for different choices of terminal distributions. In contrast, with exponential tilting the distribution of the estimator appears heavy-tailed. Consequently, we expect that 
 a large number of Monte Carlo sampling is required in order to
 obtain 
 an accurate estimation of the first moment. We 
 nevertheless expect that 
 exponential tilting could be particularly valuable when dealing with multidimensional processes, where the application of backtracking becomes challenging.
We also keep in mind that
exponential tilting is more restrictive, in the sense that it requires
 the existence of the cumulant generating function of the transition probabilities
 of the Markov process
 (as mentioned at the end of Section \ref{s34}).
 In contrast, there is no similar restriction with backtracking.
 
There is a manifest need for future research in this topic. 
In our numerical examples, we showed that there are optimal choices of the tilting parameter and of the terminal distribution, that minimize the errors of exponential tilting and backtracking, respectively. Analytical optimality results and practical formulae for the tilting parameter and for the terminal distribution 
are important questions that are mainly open. 
Another subject of investigation could
be the extension of the exponential tilting 
likelihood ratio in Eq. (\ref{e29})
from one to several tilting parameters: at each transition
of the process, a specific tilting parameter could 
be used, thus depending on the current state.
Finally, we could consider other practical settings
for comparing backtracking and exponential 
tilting, such as the one
of the insurer ruin with recuperation (cf. e.g.
\cite{gatto2015} in relation with spectrally negative L\'evy processes).

 \section{Code availability}
 The Python codes used for the creation of the figures of this article are freely available at~\cite{Aguilar2024},
 with appropriate credit to the authors.
 
\printbibliography
\clearpage
\newpage
\clearpage

 \setcounter{page}{1}
 \begin{center}
 \Large{\bf  Supplemental Material}
 \end{center}
 \appendix

\renewcommand{\thesection}{SM\arabic{section}} 
\renewcommand{\theequation}{SM\arabic{equation}}
\renewcommand{\thefigure}{SM\arabic{figure}}
\setcounter{equation}{0}

\section{$\quad$Sampling algorithms}                           \label{algo}

This section provides the general structure of the important Monte Carlo 
algorithms introduced in this article.
\begin{tcolorbox}
\begin{algorithm}[Exponential tilting for partial sums of i.i.d. random variables; Section \ref{s33}]\label{al020} \hspace{1mm}

Select a large number $m$ of iterations.
\begin{enumerate}
\item Repeat for $j=1,\ldots,m$,
\begin{itemize}
\item
generate $Y_1^{(j)},\ldots,Y_{t^\dag}^{(j)}$ independently from the d.f. $F_\theta$, cf. Eq.
(\ref{e88}), and compute $X_{t}^{(j)}=\sum_{t=1}^{t} Y_i^{(j)}$,
for $t=1,\ldots,t^\dag$;
\item Obtain $Z_{t^\dag}^{(j)}(\theta)$,
the $j$-th replication of the estimator given by Eq. (\ref{e500}).
\end{itemize}
\item Compute the importance sampling 
estimator of $z_{t^\dag}$, cf. Eq. (\ref{e79}),
given by
$$\hat{z}_{t^\dag,m} = m^{-1} \sum_{j=1}^m 
Z_{t^\dag}^{(j)}(\theta).$$
\end{enumerate}
\end{algorithm}
\end{tcolorbox}
\begin{tcolorbox}
\begin{algorithm}[Exponential tilting for Markov process; Section \ref{s34}]\label{al:tilting}
$ $\\
\hspace*{3truemm} Select a large number $m$ of iterations and fix the
initial state $X_0=n_0$.
\begin{enumerate}
    \item Repeat for $j=1,\ldots,m$, 
    \begin{itemize}    
    \item generate $X_1^{(j)},\ldots, X_{t^\dag}^{(j)}$,
 the $j$-th sample path of the Markov process, recursively
    from the exponentially tilted transition probabilities given by Eq.~\eqref{e38} and Eq.~\eqref{e40};
    \item
    obtain $Z_{t^\dag}^{(j)}(\theta)$,
    the $j$-th replication of the estimator given by Eq. (\ref{e344}).
    \end{itemize}
    \item Compute the estimator of $z_{t^\dag}$, 
    cf. Eq.~\eqref{e129}, given by
    \begin{align*}
        \hat{z}_{t^\dag,m} 
        & =\frac{1}{m}\sum_{j=1}^m Z_{t^\dag}^{(j)}(\theta). 
    \end{align*}
\end{enumerate}
\end{algorithm}
\end{tcolorbox}
\begin{tcolorbox}
\begin{algorithm}[Backtracking for Markov process; Section \ref{s42}]\label{al:backtracking}
$ $\\
\hspace*{3truemm} Select a large number $m$ of iterations and fix the
initial state $n_0$.
\begin{enumerate} 
\item Repeat for $j=1,\ldots,m$, 
\begin{itemize}
\item
generate $X^{(j)}_{t^{\dag}}$ from $w_{t^{\dag}}(\cdot;n_0)$, cf. Eq. (\ref{eq:distribution_last_state_backtrack});
\item generate $X^{(j)}_{t^{\dag}-1},\dots,X^{(j)}_0$ from the backward bridge generator given by Eq.~\eqref{eq:proof_backtracking_rate}, with backward departure 
point $X^{(j)}_{t^{\dag}}$;
\item
    obtain $Z_{n_0,t^\dag}^{(j)}$,
    the $j$-th replication of the estimator given by Eq. (\ref{e948}).
\end{itemize}
\item Compute the backtracking 
estimator of $z_{t^{\dag}}$ given by
$$\hat{z}_{t^{\dag},n_0,m} = m^{-1} \sum_{j=1}^m Z^{(j)}_{n_0,t^{\dag}}.$$
\end{enumerate}
\end{algorithm}
\end{tcolorbox}
\begin{tcolorbox}
\begin{algorithm}[Exponential tilting for binomial process, Section \ref{subsec:exampleI}]                         \label{al40}
$ $ \\
Select a large number $m$ of iterations.
\begin{enumerate}
\item Repeat for $j=1,\ldots,m$, 
\begin{itemize}
    \item generate $N_{t^{\dag}}^{(j)}$ from the 
    binomial distribution
    $B(\cdot,q,t^{\dag})$ with $q=(1+\rho)/2$ and
    $\rho$ obtained from Eq.~\eqref{eq:def_rho},
    representing the number of positive jumps of the process in the time interval $[0,t^{\dag}]$, under the exponentially 
    tilted measure;
    \item transform the generated number of positive jumps into the value of the final state, given by
    \begin{equation*}
        X^{(j)}_{t^{\dag}}=2N_{t^{\dag}}^{(j)}-t^{\dag}+n_0;
    \end{equation*}
    \item obtain the likelihood ratio
    $L_{t^{\dag}}^{(j)}(\theta)$ from Eq.
    (\ref{eq:likelihood_ratio_exponential_tilt_UBRW_V2}) with $N_{t^{\dag}}^{(j)}$.
\end{itemize}
    \item Compute the estimator of $z_{t^\dag}(\I_c(a))=\P[X_{t^{\dag}}\in \I_c(a)]$ given by
    \begin{align}\label{eq:estimator_exponential_tilted_example1}
       \!\!\!\!\! \hat{z}_{t^\dag,m}(\I_c(a)) & =\frac{1}{m}\sum_{j=1}^m I_{\I_c(a)}\left(X^{(j)}_{t^{\dag}}\right)L_{t^{\dag}}^{(j)}(\theta)
        =\frac{1}{m}\sum_{j=1}^m I_{\I_c(a)}\left(X^{(j)}_{t^{\dag}}\right)\left(\frac{r}{q}\right)^{N^{(j)}_{t^{\dag}}}\left(\frac{1-r}{1-q}\right)^{t^{\dag}-N^{(j)}_{t^{\dag}}}.
    \end{align}
\end{enumerate}
\end{algorithm}
\end{tcolorbox}
\begin{tcolorbox}
\begin{algorithm}[Backtracking for 
binomial process; Section \ref{subsec:exampleI}]   \label{al50}
$ $ \\
Select a large number $m$ of iterations.
\begin{enumerate}
\item Repeat for $j=1,\ldots,m$, 
\begin{itemize}
    \item 
    generate $N_{t^{\dag}}^{(j)}$ from $f_{t^{\dag}}$,
    representing the number of positive jumps of the process in the time interval $[0,t^{\dag}]$, under the backtracking measure;
    \item transform this number of positive jumps into 
    the value of the final state, given by 
    \begin{equation*}
        X^{(j)}_{t^{\dag}}=2N_{t^{\dag}}^{(j)}-t^{\dag}+n_0;
    \end{equation*}
    \item obtain the likelihood ratio $L_{t^{\dag}}^{(j)}(n_0)$ from (\ref{eq:likelihood_ratio_backtrack_UBRW}) with
    $X^{(j)}_{t^{\dag}}$ or with $N^{(j)}_{t^{\dag}}$.
    \end{itemize}
\item Compute the estimator of ${z}_{t^\dag}(\I_c(a))=\P[X_{t^{\dag}}\in \I_c(a)]$ given by
    \begin{equation*} \!\!\!\!
        \hat{z}_{t^\dag,m}(\I_c(a)) =\frac{1}{m}\sum_{j=1}^m I_{\I_c(a)}\left(X^{(j)}_{t^{\dag}}\right)L_{t^{\dag}}^{(j)}(n_0) =\frac{2D}{m}\sum_{j=1}^m I_{\I_c(a)}\left(X^{(j)}_{t^{\dag}}\right)B(N_{t^{\dag}}^{(j)},r,t^{\dag}),
    \end{equation*}
    where $B$ is the binomial probability function in Eq. (\ref{e184}).
\end{enumerate}
\end{algorithm}
\end{tcolorbox}

\section{$\quad$Conditioned moments of binomial process}\label{ap:conditioned_moments_binomial}

This section provides the detailed derivation of the first two moments of the binomial bridge, i.e. the binomial process $\{X_t\}_{t \in [0,t^\dag]}$ conditioned to both endpoints
$X_0=a \in \Z$ and $X_{t^\dag}=b \in \Z$. These moments
are used  in Section~\ref{sec:Conditioned_MP}.
Consider $0\le s \le t \le {t^{\dag}}$.
For any Markov process $\{X_t\}_{t \in [0,t^\dag]}$
and for any random variable $Z_t=f(X_t)$, for some
$f\!: \Z \to \Z$, its the conditional
expectation can be computed as follows,
\begin{align}\label{eq:conditioned_average}
    \E\left[Z_{t}|X_{t^{\dag}}=b,X_0=a\right] &=\sum_{j=-\infty}^\infty j \P[Z_t = j |X_{t^{\dag}}=b,X_0=a]
    = \sum_{j=-\infty}^\infty j \frac{\P[Z_t=j,X_{t^{\dag}}=b,X_0=a]}{\P[X_{t^{\dag}}=b,X_0=a]}\nonumber \\
    &= \sum_{j=-\infty}^\infty j\frac{\P[X_{t^{\dag}}=b|Z_t=j,X_0=a]\P[Z_t=j|X_0=a]}{\P[X_{t^{\dag}}=b|X_0=a]}\nonumber \\
    &= \sum_{j=-\infty}^\infty j \frac{\P[X_{t^{\dag}}=b|Z_t=j]\P[Z_t=j|X_0=a]}{\P[X_{t^{\dag}}=b|X_0=a]}, 
\end{align}
by using the Markov property. 

When $\{X_t\}_{t \in [0,t^\dag]}$ is a binomial process (cf. Section~\ref{sec:homogeneous_binomial_process}) and
for $n,n' \in \Z$,
the conditional probabilities are
\begin{equation}\label{eq:binomial_jump}
\P[X_t=n'|X_s=n]=
\binom{t-s}{\frac{n'-n+t-s}{2}}
r^{\frac{n'-n+t-s}{2}}(1-r)^{\frac{n'-n-t+s}{2}}.
\end{equation}
By inserting Eq.~\eqref{eq:binomial_jump} into Eq.~\eqref{eq:conditioned_average}, one obtains
\begin{align*}
    \E\left[Z_{t}|X_{t^{\dag}},X_0\right] =\binom{{t^{\dag}}}{\frac{X_{t^{\dag}}+t}{2}}^{-1} \sum_{n=0}^t f(n)\binom{{t^{\dag}}-t}{\frac{X_{t^{\dag}}-n+t}{2}}\binom{t}{\frac{n+t}{2}}.
\end{align*}
We can use the above formula for the case $Z_t=X_t$ and $Z_t=X_t^2$, thus yielding the first two moments
\begin{align*}
    \E\left[X_{t}|X_{t^{\dag}},X_0=0\right] & =\frac{t}{{t^{\dag}}}X_{t^{\dag}} \; \text{ and } \\
    \E\left[X_{t}^2|X_{t^{\dag}},X_0=0\right] & =  (X_{t^{\dag}}+{t^{\dag}})t\frac{(X_{t^{\dag}}+{t^{\dag}})(t-1)-2t+2{t^{\dag}}}{{t^{\dag}}({t^{\dag}}-1)}
    +2t^2-\frac{t^2(X_{t^{\dag}}+{t^{\dag}})}{{t^{\dag}}}.
\end{align*}
Mean and variance can thus be computed. They are respectively
given by
\begin{align}                               \label{e1000}
\mu_t = \E\left[X_{t}|X_{t^{\dag}},X_0=0\right] 
\text{ and } \sigma_t^2 = \E\left[X_{t}^2|X_{t^{\dag}},X_0=0\right]
- \E^2\left[X_{t}|X_{t^{\dag}},X_0=0\right].
\end{align}
\end{document}